\documentclass[journal]{IEEEtran}
\makeatletter
\def\ps@headings{%
\def\@oddhead{\mbox{}\scriptsize\rightmark \hfil \thepage}%
\def\@evenhead{\scriptsize\thepage \hfil \leftmark\mbox{}}%
\def\@oddfoot{}%
\def\@evenfoot{}}
\makeatother
\usepackage{subfigure}
\usepackage{colortbl}
\newcommand{\bm}[1]{\mbox{\boldmath{$#1$}}}

\usepackage{graphicx}  
\usepackage{url}       
\usepackage{bm}

\usepackage{amsmath}   
\usepackage{cite}
\usepackage{amsfonts,amssymb}

\usepackage{stfloats}

\usepackage{cases}
\usepackage{algorithm}
\usepackage{multirow}
\usepackage{algorithmic}
\usepackage{stfloats}
\usepackage{epstopdf}
\makeatletter

\newcommand{\Rmnum}[1]{\expandafter\@slowromancap\romannumeral #1@}
\makeatother

\newtheorem{proof}{Proof}

\newcommand{\ls}[1]
    {\dimen0=\fontdimen6\the\font
     \lineskip=#1\dimen0
     \advance\lineskip.5\fontdimen5\the\font
     \advance\lineskip-\dimen0
     \lineskiplimit=.9\lineskip
     \baselineskip=\lineskip
     \advance\baselineskip\dimen0
     \normallineskip\lineskip
     \normallineskiplimit\lineskiplimit
     \normalbaselineskip\baselineskip
     \ignorespaces
    }

\begin{document}
\title{
Double-RIS-Assisted Orbital Angular Momentum Near-Field Secure Communications
}
\author{\IEEEauthorblockN{Liping Liang, \emph{Member, IEEE}, Minmin Wang, \emph{Student Member, IEEE}, Wenchi Cheng, \emph{Senior Member, IEEE}, \\
and Wei Zhang, \emph{Fellow, IEEE}}\\[0.2cm]




\thanks{Liping Liang, Minmin Wang, and Wenchi Cheng are with the State Key Laboratory of Integrated Services Networks, Xidian University, Xi'an 710071, China (e-mail: liangliping@xidian.edu.cn; minminwang@stu.xidian.edu.cn; wccheng@xidian.edu.cn).

Wei Zhang is with the School of Electrical Engineering and Telecommunications, The University of New South Wales, Sydney, NSW 2052, Australia
(e-mail: w.zhang@unsw.edu.au).
	
}
}

\maketitle

\begin{abstract}

To satisfy the various demands of growing devices and services, emerging high-frequency-based technologies promote near-field wireless communications. Therefore, near-field physical layer security has attracted much attention to facilitate the wireless information security against illegitimate eavesdropping. However, highly correlated channels between legitimate transceivers and eavesdroppers of existing multiple-input multiple-output (MIMO) based near-field secure technologies along with the low degrees of freedom significantly limit the enhancement of security results in wireless communications. To significantly increase the secrecy rates of near-field wireless communications, in this paper we propose the double-reconfigurable-intelligent-surface (RIS) assisted orbital angular momentum (OAM) secure scheme, where RISs with few reflecting elements are easily deployed to reconstruct the direct links blocked by obstacles between the legitimate transceivers, mitigate the inter-mode interference caused by the misalignment of legitimate transceivers, and adjust the OAM beams direction to interfere with eavesdroppers. Meanwhile, due to the unique orthogonality among OAM modes, the OAM-based joint index modulation and artificial noise scheme is proposed to weaken the information acquisition by eavesdroppers while increasing the achievable rate with the low cost of legitimate communications. To maximize the secrecy rate of our proposed scheme, we develop the Riemannian manifold conjugate gradient (RMCG)-based alternative optimization (AO) algorithm to jointly optimize the transmit power allocation of OAM modes and phase shifts of double RISs. Numerical results show that our proposed double-RIS-assisted OAM near-field secure scheme outperforms the existing works in terms of the secrecy rate and the eavesdropper's bit error rate.

\end{abstract}


\begin{IEEEkeywords}
	Orbital angular momentum (OAM), near-field secure communications, reconfigurable intelligent surface (RIS),  physical layer security, secrecy rate.
\end{IEEEkeywords}


\section{Introduction}

\IEEEPARstart{T}{o} satisfy the growing demands of tremendous devices and services in the future wireless communications, emerging technologies, such as millimeter wave (mmWave), terahertz (THz) and orbital angular momentum (OAM), recently have been investigated by academics~\cite{9411894,9398864,8558508}. These above-mentioned technologies rely on the utilization of high frequency and helical phases which cause severe propagation attenuation and divergence, respectively, in long-distance wireless communications. Thus, they generally implement the performance enhancement of wireless communications with the assistance of resource allocation, interference management, beam steering, and other methods in the near field.

Due to the broadcast and open characteristics of wireless channels, legitimate communications are vulnerable to being intercepted by malicious eavesdroppers, thus severely degrading the security results. Therefore, physical layer security (PLS) has become a key research area to facilitate the wireless information security against illegitimate eavesdropping with high-reliability information transmission~\cite{9952192,8353854,9844767}. To achieve the expected anti-eavesdropping results of wireless communications, existing literature on PLS, such as multiple-input multiple-output (MIMO) systems aided by joint transmit and receive beamforming, artificial noise (AN), and large-scale reconfigurable intelligent surface (RIS)~\cite{Beamforming_Power_Design_for_IRS,IRS_assisted_untrusted_NOMA,AN_Aided_Secure_MIMO_via_IRS}, mainly focuses on exploiting the physical characteristics of received signals while fully employing the degrees of freedom (DoFs) of MIMO without relying on very complicated secret key generation and sophisticated encryption systems.

The adaptive PLS algorithm was proposed to secure both data and pilots under correlated and uncorrelated eavesdropping channels in orthogonal-frequency-division-multiplexing (OFDM) based MIMO systems~\cite{9713869}. The nonorthogonal multiple access (NOMA) and RIS technologies are integrated to bring more gain in both energy efficiency and secrecy performance for wireless communications than traditional NOMA technologies~\cite{9385957,9760141,10057422}. The authors of~\cite{9491934} have designed the noniterative secure transceivers and power allocation scheme to increase the secrecy rate for AN-assisted MIMO networks. Also, joint precoding and AN design were investigated to maximize the sum secrecy rate for multi-user-based MIMO wiretap channels with multiple eavesdroppers~\cite{9973364}. Nevertheless, the highly correlated channels between eavesdroppers and legitimate transceivers in line-of-sight (LoS) MIMO systems make the difficult to efficiently distinguish eavesdropping channels and legitimate channels in the angular domain with conventional plane electromagnetic waves. Meanwhile, LoS MIMO has low DoFs. 
Thus, it is very difficult to flexibly apply existing MIMO-based secure systems to significantly improve the security in LoS scenarios, which highly prompts the exploration of novel technologies to significantly enhance the PLS of near-field wireless communications.


As mentioned above, OAM, which is associated with the helical phase fronts of electromagnetic waves, shows great advantages to significantly increase the spectrum efficiency and enhance the PLS of near-field wireless communications~\cite{8894467,8960405,10232879,9690469,8002571,8712342,9046265}. The main reasons are given as follows: 1) Multiplexing signals can be transmitted in parallel among multiple OAM modes due to the inherent orthogonality among different integer OAM modes with the perfectly aligned transceivers. Several studies have been devoted to addressing the challenges of misalignment of transceivers in most practical communication scenarios~\cite{8761299,9714507,10239524}. The negative impact of OAM mode offset on the spectrum efficiencies of point-to-point LoS OAM systems was theoretically analyzed~\cite{9543535}. The uniform circular array (UCA)-based THz multi-user OAM communication scheme was proposed, which consists of downlink and uplink transmission methods, to achieve the same spectrum efficiency but with much lower implementation complexity as compared with traditional MIMO scheme~\cite{10040606,7797488}. 2) The received OAM signals by eavesdroppers are sensitive to helical phases and the OAM beams are central hollow, thus making it difficult for eavesdroppers to acquire legitimate information. The authors of~\cite{9844288} have proposed the physical layer secret key generation scheme, where a mode-based active defense phase is designed to invalidate keys generated by eavesdroppers, to reach a high key generate rate in OAM secure communications. The frequency-diverse-array-based direction-range-time dependent OAM directional modulation scheme was proposed to improve the security performance with no AN~\cite{9749285}.
However, the existing works on OAM secure communications primarily focus on designing antenna arrays or secret key generation while the strategies for exploiting the physical characteristics of wireless channels in the near field still lack studies. We primarily studied the passive RIS-assisted OAM for secure wireless communications under the aligned legitimate transceivers, where the LoS links are not blocked by obstacles, to verify the superiority of OAM in anti-eavesdropping~\cite{sub}.  

Motivated by the above-mentioned problem, in this paper we propose the double-RIS-assisted OAM secure near-field communication scheme, where OAM is employed to achieve high DoFs and simple implementation complexity with the inverse discrete Fourier transform (IDFT) operation at the legitimate transmitter and the discrete Fourier transform (DFT) operation at the legitimate receiver, to significantly increase the secrecy rates of near-field wireless communications. In the proposed scheme, we assuthe direct links between the UCA-based misaligned legitimate transceivers are blocked by obstacles. Hence, RISs are easily deployed to mitigate the inter-mode interference caused by the misalignment of legitimate transceivers, reconstruct the direct links blocked by obstacles between the legitimate transceivers, and adjust the OAM beams direction to interfere with eavesdroppers. Meanwhile, the OAM-based joint index modulation and artificial noise (JiMa) scheme is proposed to weaken the information acquisition by eavesdroppers while increasing the achievable rate with the low cost of legitimate communications. Then, the secrecy rate maximization problem is formulated to improve the anti-eavesdropping performance of near-field wireless communications by jointly optimizing the transmit power allocation of OAM modes and phase shifts of double RISs. To solve the non-convex optimization problem with intricately coupled optimization variables, we develop the Riemannian manifold conjugate gradient (RMCG) based alternative optimization (AO) algorithm. Extensive numerical results have shown that our proposed double-RIS-assisted OAM near-field secure scheme outperforms the existing works in terms of the secrecy rate and the eavesdropper's bit error rate.

The remainder of this paper is organized as follows. Section~\ref{sec:sys} presents the system model of our proposed double-RIS-assisted OAM secure near-field communications. Then, we propose the scheme in Section~\ref{sec:scheme}, which includes the signal processing and secrecy rate optimization problem formulation. Section~\ref{sec:Algorithm} develops the RMCG-AO algorithm to achieve the maximum secrecy rate and Section~\ref{sec:Results} shows the numerical results to evaluate the performance of our proposed scheme. Finally, Section~\ref{sec:Conclusion} concludes this paper.

{\emph Notations}: $[\cdot]^{\rm T}$ and $(\cdot)^{\rm H}$ represent the transpose and conjugate transpose of a matrix, respectively. Boldface lowercase and uppercase letters respectively denote vectors and matrices. $\lfloor\cdot\rfloor$ denotes the floor function and ${a\choose b}$ represents the combinations by choosing $b$ from $a$. Also, $|\cdot|$, $\mathbb{E}(\cdot)$, $\parallel\cdot\parallel$, and $\operatorname{diag}\left(\cdot\right)$ represent the absolute value of a scalar, the expectation operation, the Euclidean norm of a vector, and the diagonal matrix with a vector on the main diagonal, respectively. $\mathbf{n}_\mathrm{a} \sim \mathcal{C N}\left(0, \sigma_\mathrm{a}^{2} \mathbf{I}_{N}\right)$ is the Gaussian noise with zero mean and variance $\sigma_\mathrm{a}^{2}$, where $\mathbf{I}_{N}$ denotes the $N \times N$ identity matrix. In addition, $(\cdot)^{-1}$, $\left( \cdot\right)^{*}$, $\circ$, and $\Re \left\{\cdot\right\}$ denote the inverse of a matrix or a scalar, the conjugate of a matrix or vector, the Hadamard product operation, and the real part of a complex number, respectively.

\begin{figure*}[htbp]
	\centering
	\includegraphics[width=1.01\textwidth]{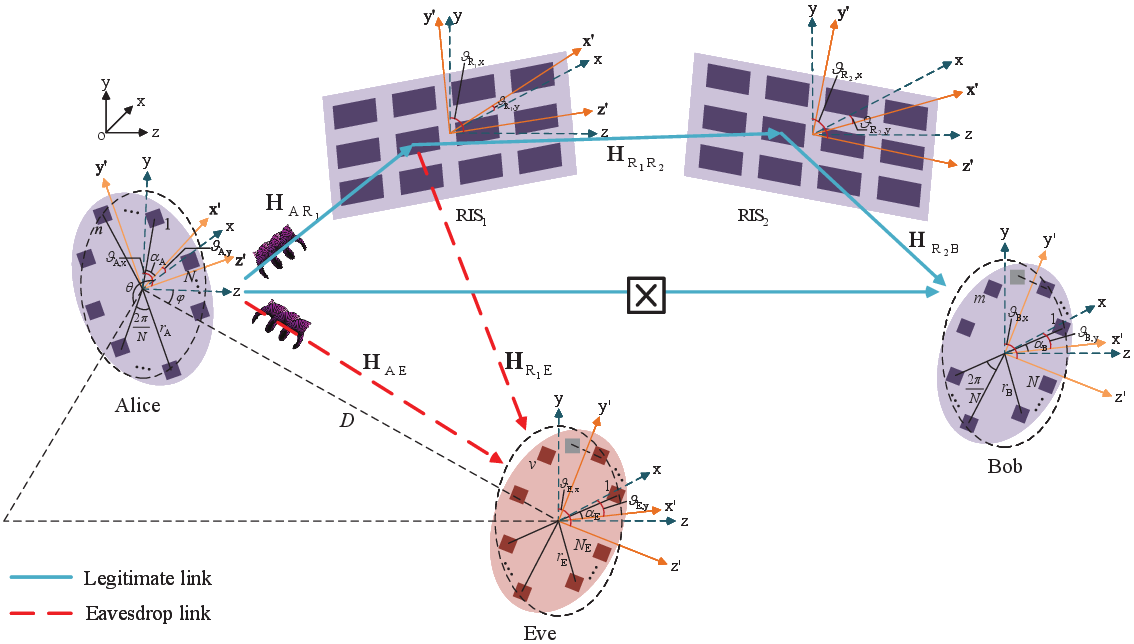}
	\caption{The system model of our proposed double-RIS-assisted OAM secure near-field communications.} \label{fig: system model}
\end{figure*}

\section{System Model}\label{sec:sys}

To generate and receive multiple OAM signals, the legitimate transmitter (called Alice) and the legitimate receiver (called Bob) are equipped with $N$ arrays equidistantly distributed around the circles with the radii $r_{\rm A}$ and $r_{\rm B}$, respectively. The phase difference between any two adjacent arrays is $e^{j\frac{2\pi\l}{N}}$ for Alice, where $ l$ $\left(  0\leq l \leq N-1\right) $ represents the order of generated OAM modes. In some practical near-field wireless communication scenarios, the direct link between Alice and Bob is blocked by obstacles. An eavesdropper (called Eve) with a $N_{\rm E}$-array UCA attempts to intercept the legitimate OAM wireless communications. To implement the secure wireless communications between Alice and Bob, we build a double-RIS-assisted OAM secure near-field communication system as depicted in Fig.~\ref{fig: system model}, where Alice and Bob are misaligned. Aiming at mitigating the inter-mode interference caused by the misalignment between Alice and Bob, strengthening the interference to the received signals by Eve, and establishing alternative direct communication links from Alice to Bob, double rectangular RISs for easy deployment and the low cost are utilized to flexibly adjust the OAM beams direction in the proposed system. The passive RIS close to Alice, denoted by RIS$_{1}$, is equipped with $Q_{1}=Q_{\mathrm{y,1}} Q_{\mathrm{z,1}}$ reflecting elements, where $ Q_{\mathrm{y,1}} $ and $ Q_{\mathrm{z,1}} $ are the number of reflecting elements along the y-axis and z-axis, respectively. $Q_{2}=Q_{\mathrm{y,2}} Q_{\mathrm{z,2}}$ reflecting elements are integrated on the other active RIS close to Bob, denoted by RIS$_{2}$, where $ Q_{\mathrm{y,2}} $ and $ Q_{\mathrm{z,2}} $ respectively represent the number of reflecting elements along the y-axis and z-axis. If RIS$_{1}$ is completely overlapped with RIS$_{2}$ at the same location, the inter-mode interference cannot be effectively mitigated in our proposed systems. Thus, RIS$_{1}$ and RIS$_{2}$ are deployed in different locations.

As shown in Fig.~\ref{fig: system model}, we utilize the Cartesian coordinate system with the x-axis, y-axis, and z-axis to clearly explain the locations of RISs, Alice, Bob, and Eve in the double-RIS-assisted OAM secure near-field communication system. Also, to express the coordinates of each array on Alice, Bob, Eve, and RISs, we can obtain the virtual coordinate system x$^{\prime}$y$^{\prime}$z$^{\prime}$ formed by rotating around the x-axis, y-axis, and z-axis. First, we give the coordinate of each array on Alice and Bob. The center of Alice, represented with $\mathbf{u}_{\mathrm{A}}=[0,0,0]^{\mathrm{T}}$, is located at the origin of coordinates. The coordinate corresponding to the center of Bob is denoted by $\mathbf{u}_{\mathrm{B}}=[x_{\mathrm{B}},y_{\mathrm{B}} ,z_{\mathrm{B}} ]^{\mathrm{T}}$, where $ x_\mathrm{B} $, $ y_\mathrm{B} $, and $ z_\mathrm{B} $ represent the coordinates of the x-axis, y-axis, and z-axis. We denote by $\alpha_{\mathrm{A}}$ the initial azimuth angle between the first array on Alice and the x-axis. Thus, we have the azimuth angle for the $n$-th $(1\leq n\leq N)$ array on Alice as $\phi_{n}=\frac{2 \pi(n-1)}{N}+\alpha_{\mathrm{A}}$. Also, we denote by $\vartheta_{\mathrm{A,x}}$ and $\vartheta_{\mathrm{A,y}}$ the rotation angles around the x-axis and y-axis for Alice, respectively. $\mathbf{R}_{\mathrm{A,x}}\left(\vartheta_{\mathrm{A,x}}\right)$ and $\mathbf{R}_{\mathrm{A,y}}\left(\vartheta_{ \mathrm{A,y}}\right)$ are respectively the attitude matrices~\cite{8761299} corresponding to the x-axis and y-axis. Similarly, $\psi_{m}=\frac{2 \pi(m-1)}{N}+\alpha_{\mathrm{B}}$ is denoted by the azimuth angle for the $m$-th $(1\leq m\leq N)$ array on Bob, where $\alpha_{\mathrm{B}}$ is the initial azimuth angle between the first array on Bob and the x-axis. $\mathbf{R}_{\mathrm{B,x}}\left(\vartheta_{\mathrm{B,x}}\right)$ and $\mathbf{R}_{\mathrm{B,y}}\left(\vartheta_{ \mathrm{B,y}}\right)$ are respectively the attitude matrices corresponding to the x-axis and y-axis for Bob, where $\vartheta_{\mathrm{B,x}}$ and $ \vartheta_{\mathrm{B,y}} $ represent the rotation angles around the x-axis and y-axis, respectively.

Therefore, the coordinate of $n$-th array on Alice and $m$-th array on Bob is respectively expressed as follows:
\begin{equation}
 \mathbf{u}_{\mathrm{A}, n}= \mathbf{u}_{\mathrm{A}}+\mathbf{R}_{ \mathrm{A}}\left[r_{\rm A} \cos \left( \phi_{n}\right), r_{\rm A} \sin \left( \phi_{n}\right) ,0 \right]^{\mathrm{T}}	
\end{equation}
and
\begin{equation}
 \mathbf{u}_{\mathrm{B}, m}= \mathbf{u}_{\mathrm{B}}+\mathbf{R}_{ \mathrm{B}}\left[r_{\rm B} \cos \left( \psi_{m}\right) , r_{\rm B} \sin \left( \psi_{m}\right) ,0 \right]^{\mathrm{T}},	
\end{equation}
where $\mathbf{R}_{ \mathrm{A}}$ and $\mathbf{R}_{ \mathrm{B}}$ are given by
\begin{equation}
\begin{cases}
\mathbf{R}_{ \mathrm{A}}=\mathbf{R}_{ \mathrm{A,y}}\left(\vartheta_{ \mathrm{A,y}}\right)\mathbf{R}_{ \mathrm{A,x}}\left(\vartheta_{ \mathrm{A,x}}\right);\\
\mathbf{R}_{\mathrm{B}}=\mathbf{R}_{\mathrm{B,y}}\left(\vartheta_{ \mathrm{B,y}}\right)\mathbf{R}_{ \mathrm{B,x}}\left(\vartheta_{ \mathrm{B,x}}\right).
\end{cases}
\end{equation}

Next, the location of Eve is given. We denote by $D$ the distance between the centers of Alice and Eve, $\varphi$ the included angle between the z-axis and the line from the center of Alice to the center of Eve, and $\theta$ the included angle between the x-axis and the projection of the line from the center of Alice to the center of Eve on the xoy plane. Thus, the center of Eve is located at $\mathbf{u}_{\mathrm{E}}=\left[D\sin\varphi\cos\theta,D\sin\varphi\sin\theta,D\cos\varphi\right]^{\mathrm{T}}$. Also, we denote by $\alpha_{\mathrm{E}}$ the initial azimuth angle between the first array on Eve and the x-axis, $\mathbf{R}_{\mathrm{E,x}}\left(\vartheta_{\mathrm{E,x}}\right)$ the attitude matrix with respect to the rotation angle $\vartheta_{\mathrm{E,x}}$ around the x-axis for the Eve, and $\mathbf{R}_{\mathrm{E,y}}\left(\vartheta_{ \mathrm{E,y}}\right)$ the attitude matrix with respect to the rotation angle $\vartheta_{\mathrm{E,y}}$ around the y-axis for the Eve. Thereby, we have the coordinate of the $v$-th $ (1\leq v\leq N_{\rm E}) $ array, denoted by $\mathbf{u}_{\mathrm{E}, v}$, on Eve as follows:
\begin{equation}
 \mathbf{u}_{\mathrm{E}, v}= \mathbf{u}_{\mathrm{E}}+\mathbf{R}_{\mathrm{E}}\left[ r_{\mathrm{E}} \cos \left(\kappa_{v}\right),r_{\mathrm{E}} \sin \left(\kappa_{ v}\right),0\right]^{\mathrm{T}},	
\end{equation}
where $\kappa_{v}=\frac{2 \pi(v-1)}{N_{\rm E}}+\alpha_{\mathrm{E}}$ is the azimuth angle for the $v$-th array on Eve, $r_{\mathrm{E}}$ represents the radius of Eve, and $\mathbf{R}_{\mathrm{E}}=\mathbf{R}_{\mathrm{E,y}}\left(\vartheta_{ \mathrm{E,y}}\right)\mathbf{R}_{ \mathrm{E,x}}\left(\vartheta_{ \mathrm{E,x}}\right)$.

To reconstruct the direct links from Alice to Bob, we then construct the location expressions of RISs in our proposed double-RIS-assisted OAM secure communications system. As shown in Fig.~\ref{fig: system model}, we have the coordinates with respect to the center points of RIS$_{1}$ and RIS$_{2}$ as $\mathbf{u}_{\mathrm{R}_{1}}=\left[x_{\mathrm{R}_{1}}, y_{\mathrm{R}_{1}}, z_{\mathrm{R}_{1}}\right]^{\mathrm{T}}$ and $\mathbf{u}_{\mathrm{R}_{2}}=\left[x_{\mathrm{R}_{2}}, y_{\mathrm{R}_{2}}, z_{\mathrm{R}_{2}}\right]^{\mathrm{T}}$, where $x_{\mathrm{R}_{1}} $ (or $x_{\mathrm{R}_{2}}$), $y_{\mathrm{R}_{1}} $ (or $y_{\mathrm{R}_{2}}$), and $z_{\mathrm{R}_{1}} $ (or $z_{\mathrm{R}_{2}}$) are the coordinates along the x-axis, y-axis, and z-axis, respectively, for the RIS$_{1}$ (or RIS$_{2}$).

Based on $\mathbf{u}_{\mathrm{R},1}$ and $\mathbf{u}_{\mathrm{R},2}$, the coordinates of the $q_{1}$-th $(1\leq q_{1}\leq Q_{1})$ reflecting element on the RIS$_{1}$ and the $q_{2}$-th $(1\leq q_{2}\leq Q_{2})$ reflecting element on the RIS$_{2}$, denoted by $\mathbf{u}_{\mathrm{R}_{1}, q_{1}}$ and $\mathbf{u}_{\mathrm{R}_{2}, q_{2}}$, can be respectively expressed by
\begin{equation}
\begin{cases}
   \hspace{-1mm} \mathbf{u}_{\mathrm{R}_{1}, q_{1}} \hspace{-1mm} =\hspace{-1mm} \mathbf{u}_{\mathrm{R}_{1}}\hspace{-1mm}+\hspace{-1mm}\mathbf{R}_{\mathrm{R}_{1}}\hspace{-1mm}\left[ 0,d_{\mathrm{y}}\hspace{-0.8mm}\left(\hspace{-0.5mm}q_{\mathrm{y}_{1}}\hspace{-1mm}-\hspace{-1mm}\frac{1+Q_{\mathrm{y,\hspace{-0.2mm} 1}}}{2}\hspace{-1mm}\right)\hspace{-0.5mm},
    d_{\mathrm{z}}\hspace{-0.8mm}\left(\hspace{-0.5mm}q_{\mathrm{z}_{1}}\hspace{-1mm}-\hspace{-1mm}\frac{1+Q_{\mathrm{z,\hspace{-0.2mm} 1}}}{2}\hspace{-0.5mm}\right)\hspace{-0.8mm}\right]^{\mathrm{T}}\hspace{-1mm},\\
	\hspace{-1mm}\mathbf{u}_{\mathrm{R}_{2}, q_{2}}\hspace{-1mm}  =\hspace{-1mm} \mathbf{u}_{\mathrm{R}_{2}}\hspace{-1mm}+\hspace{-1mm}\mathbf{R}_{\mathrm{R}_{2}}\hspace{-1mm}\left[ 0,d_{\mathrm{y}}\hspace{-0.8mm}\left(\hspace{-0.5mm}q_{\mathrm{y}_{2}}\hspace{-1mm}-\hspace{-1mm}\frac{1+Q_{\mathrm{y,\hspace{-0.2mm} 2}}}{2}\hspace{-1mm}\right)\hspace{-0.5mm},
d_{\mathrm{z}}\hspace{-0.8mm}\left(\hspace{-0.5mm}q_{\mathrm{z}_{2}}\hspace{-1mm}-\hspace{-1mm}\frac{1+Q_{\mathrm{z,\hspace{-0.2mm} 2}}}{2}\hspace{-0.5mm}\right)\hspace{-0.8mm}\right]^{\mathrm{T}}\hspace{-1mm},
\end{cases}
\label{eq:u_R1R2}
\end{equation}
where $q_{\mathrm{y}_{1}} \in \left\{1,2, \ldots, Q_{\mathrm{y,1}}\right\}$, $q_{\mathrm{z}_{1}} \in \left\{1,2, \ldots, Q_{\mathrm{z,1}}\right\}$, $q_{\mathrm{y}_{2}} \in \left\{1,2, \ldots, Q_{\mathrm{y,2}}\right\}$, $q_{\mathrm{z}_{2}} \in \left\{1,2, \ldots, Q_{\mathrm{z,2}}\right\}$, and $d_{\mathrm{y}}$ as well as $d_{\mathrm{z}}$ are respectively the element separation distances on the RIS$_{1}$ (RIS$_{2}$) along the y$'$-axis and z$'$-axis. In Eq.~\eqref{eq:u_R1R2}, we have
 \begin{equation}
\begin{cases}
\mathbf{R}_{\mathrm{R}_{1}}=\mathbf{R}_{\mathrm{R_{1},y}}\left(\vartheta_{ \mathrm{R_{1},y}}\right)\mathbf{R}_{ \mathrm{R_{1},x}}\left(\vartheta_{ \mathrm{R_{1},x}}\right);\\
\mathbf{R}_{\mathrm{R}_{2}}=\mathbf{R}_{\mathrm{R_{2},y}}\left(\vartheta_{ \mathrm{R_{2},y}}\right)\mathbf{R}_{ \mathrm{R_{2},x}}\left(\vartheta_{ \mathrm{R_{2},x}}\right),
\end{cases}
\end{equation}
where $\mathbf{R}_{\mathrm{R_{1},x}}\left(\vartheta_{\mathrm{R_{1},x}}\right)$ and $\mathbf{R}_{\mathrm{R_{1},y}}\left(\vartheta_{ \mathrm{R_{1},y}}\right)$ are respectively the attitude matrices with respect to the rotation angles $\vartheta_{\mathrm{R_{1},x}}$ and $\vartheta_{ \mathrm{R_{1},y}}$ around  the x-axis and y-axis for the RIS$_{1}$, $\mathbf{R}_{\mathrm{R_{2},x}}\left(\vartheta_{\mathrm{R_{2},x}}\right)$ and $\mathbf{R}_{\mathrm{R_{2},y}}\left(\vartheta_{ \mathrm{R_{2},y}}\right)$ are respectively the attitude matrices corresponding to the rotation angles $\vartheta_{\mathrm{R_{2},x}}$ and $\vartheta_{ \mathrm{R_{2},y}}$ around  the x-axis and y-axis for the RIS$_{2}$.

Based on the proposed secure communications system and inspired by index modulation, we propose the double-RIS-assisted OAM with JiMa scheme to significantly enhance the secrecy results in the near field.

\section{The Double-RIS-Assisted OAM with JiMa Scheme for High Secrecy Rate}\label{sec:scheme}
\subsection{Signal Processing}

To improve the secrecy performance of our proposed double-RIS-assisted OAM system, the available OAM modes at Alice are separated into one part for desired signal transmissions and the other part for carrying AN. Considering the divergence of OAM beams and the requirement of high spectrum efficiencies of near-field legitimate wireless communications, we have $N_\mathrm{A}$ low-order OAM modes including zero OAM mode for desired signal transmissions and the remaining $\left( N-N_\mathrm{A}\right)$ high-order OAM modes for sending AN.

Aiming at decreasing the cost and further increasing the spectrum efficiency of our proposed double-RIS-assisted OAM system, the index modulation is leveraged to divide the input signals at Alice into modulated desired signal as well as AN information and the OAM-AN index information~\cite{9690469}. According to the index information, we assume that the desired signals are emitted by $N_\mathrm{s}$ OAM modes with the set $\mathcal{L}_\mathrm{s}$ and AN is transmitted by $N_\mathrm{\ddot{z}}$ OAM modes with the set $\mathcal{L_\mathrm{\ddot{z}}}$, where $\mathcal{L}_\mathrm{s}$ and $\mathcal{L_\mathrm{\ddot{z}}}$ consist of $N_\mathrm{s}$ out of $N_\mathrm{A}$ low-order OAM modes and $N_\mathrm{\ddot{z}}$ out of the remaining $\left( N-N_\mathrm{A}\right)$ high-order OAM modes, respectively. It is noticed that $\mathcal{L}_\mathrm{s}$ has to include zero OAM mode to guarantee the spectrum efficiency of our proposed system. Thus, we can combine the sets $\mathcal{L}_\mathrm{s}$ and $\mathcal{L_\mathrm{\ddot{z}}}$ one by one for desired signals and AN transmission, thereby forming the signal and AN (SN) pairs. Each SN pair corresponding to the specific index information. Therefore, the number of total available SN pairs is $G\!=\! 2^{\left\lfloor\log_{2}\left[  \!\left(\!\!\!\!
	\begin{tiny}
		\begin{array}{c}
			N_\mathrm{A}-1 \\
			N_\mathrm{s}-1
		\end{array}
	\end{tiny} \!\!\!\!\right) \left(\!\!\!\!
	\begin{tiny}
		\begin{array}{c}
			N\!-\!N_\mathrm{A} \\
			N_\mathrm{\ddot{z}}
		\end{array}
	\end{tiny} \!\!\!\!\right)\right] \right\rfloor }$.
In our proposed scheme, the SN index information is pre-shared by Alice and Bob to reduce the signal detection error at Bob. Thus, $\mathcal{L}_\mathrm{s}$ and $\mathcal{L}_\mathrm{\ddot{z}}$ are pre-known by both Alice and Bob.

We denote by $\mathbf{s} \in \mathbb{C}^{N \times 1}$ and $\mathbf{z} \in \mathbb{C}^{N \times 1}$ the transmitted desired signals and AN vectors, respectively. All other entries in $\mathbf{s}$ are 0 except the entries selected to transmit the desired signals with $\mathbb{E}\left\lbrace \left\vert s_{l^\mathrm{s}_{\bar{n}}} \right\vert^2\right\rbrace  = p_{l^\mathrm{s}_{\bar{n}}}$, where $l^\mathrm{s}_{\bar{n}}$ is the $ \bar{n}$-th $\left( 1\leq \bar{n} \leq N_\mathrm{s}\right)$ OAM mode used to transmit the desired signals in $\mathcal{L}_\mathrm{s}$, $s_{l^\mathrm{s}_{\bar{n}}}$ is the transmit symbol, and $p_{l^\mathrm{s}_{\bar{n}}}$ is the allocated power to the OAM mode $l^\mathrm{s}_{\bar{n}}$. Also, all other entries in $\mathbf{z}$ are 0 except that the entries selected to transmit AN, denoted by $z_{l^\mathrm{z}_{\ddot{n}}}$, with the variance $\sigma_\mathrm{\ddot{z}}^{2}$ and zero mean, where $l^\mathrm{z}_{\ddot{n}}$ is the $\ddot{n}$-th $ \left( 1\leq \ddot{n} \leq N_\mathrm{\ddot{z}}\right)$ OAM mode used to transmit AN with equal power allocation in $\mathcal{L}_\mathrm{\ddot{z}}$. In our proposed scheme, $\rho$ $\left(0 < \rho < 1 \right)$ is denoted by the ration of the transmit power for the desired signals to the total transmit power $P_\mathrm{T}$. Thus, we have $\sigma_\mathrm{\ddot{z}}^2=\frac{\left( 1-\rho\right)P_\mathrm{T} }{N_\mathrm{\ddot{z}}}$.

Based on $\mathbf{s}$ and $\mathbf{z} $, the initial signal, denoted by $\tilde{\mathbf{s}}$, transmitted by Alice can be expressed as follows:
\begin{equation}
\tilde{\mathbf{s}}=\mathbf{s}+\mathbf{z}.
\label{eq:sz}
\end{equation}
Hence, we have the signals, denoted by $\mathbf{x}$, with respect to the whole OAM modes emitted by Alice for our proposed scheme as follows:
\begin{equation}
\mathbf{x} = \mathbf{F } \tilde{\mathbf{s}},
\label{eq:transmit signal}
\end{equation}
where $\mathbf{F}=\left[\mathbf{f}_{0}, \ldots, \mathbf{f}_{N-1}\right]  \mathbb{C}^{N \times N}$ is the IDFT matrix with $\mathbf{f}_{l}\hspace{-1mm}=\hspace{-1mm}\frac{1}{\sqrt{N}}\left[ e^{j l  \phi_{1} }, \ldots, e^{j l \phi_{N}}\right]^\mathrm{T}$.

As shown in Fig.~\ref{fig: system model}, the emitted signals $\mathbf{x}$ are sequential reflected by the RIS$_{1}$ and RIS$_{2}$ to adjust the directions of OAM beams received by Bob and Eve. We denote by $\mathbf{\Theta}_{1}=\operatorname{diag}\left(\boldsymbol{\theta}_{1}^\mathrm{H}\right)$ the diagonal phase shift matrix of RIS$_{1}$ with $\boldsymbol{\theta}_{1}=\left[\theta_{1,1}, \ldots, \theta_{1,Q_{1}}\right]^\mathrm{T}$ , $\mathbf{\Theta}_{2}=\operatorname{diag}\left(\boldsymbol{\theta}_{2}^\mathrm{H}\right)$ the diagonal phase shift matrix of RIS$_{2}$ with $\boldsymbol{\theta}_{2}=\left[\theta_{2,1}, \ldots, \theta_{2,Q_{2}}\right]^\mathrm{T}$, and $\mathbf{A}=\operatorname{diag}\left(\mathbf{a} \right) $ with $ \mathbf{a}=\left[a_{1}, \dots,  a_{Q_{2}}\right]^\mathrm{T} $ representing the amplifying coefficients of RIS$_{2}$.

The channels from RIS$_{1}$ to Eve is assumed to be blocked by obstacles as depicted in Fig.~\ref{fig: system model}. Also, the channels from Alice to RIS$_{1}$, RIS$_{1}$ to RIS$_{2}$, RIS$_{2}$ to Bob, Alice to Eve, and RIS$_{1}$ to Eve are denoted by $\mathbf{H}_{\mathrm{AR_{1}}} \in \mathbb{C}^{Q_{1} \times N}$, $\mathbf{H}_{\mathrm{R_{1}}\mathrm{R_{2}}} \in \mathbb{C}^{Q_{2} \times Q_{1}} $, $\mathbf{H}_{\mathrm{R_{2}B}} \in \mathbb{C}^{N \times Q_{2}}$, $\mathbf{H}_{\mathrm{AE}} \in \mathbb{C}^{N \times N_{\rm E}}$, and $\mathbf{H}_{\mathrm{R_{1}E}} \in \mathbb{C}^{N_{\rm E} \times Q_{1}}$, respectively. Specifically, denoting $\lambda$ the wavelength of the carrier, we can calculate the arbitrary entries of $\mathbf{H}_{\mathrm{AR_{1}}}$, $\mathbf{H}_{\mathrm{R_{1}}\mathrm{R_{2}}}$, $\mathbf{H}_{\mathrm{R_{2}B}}$, $\mathbf{H}_{\mathrm{AE}}$, and $\mathbf{H}_{\mathrm{R_{1}E}}$ as follows:
\begin{equation}
	\left\{
	\begin{array}{lr}
		h_{\mathrm{AR_{1}}, q_{1} n}  = \frac{\lambda \beta_{\mathrm{AR_{1}}} }{4 \pi \Vert \mathbf{u}_{\mathrm{R}_{1}, q_{1}}-\mathbf{u}_{\mathrm{A}, n}\Vert} e^{-j \frac{2 \pi }{\lambda} \Vert \mathbf{u}_{\mathrm{R}_{1}, q_{1}}-\mathbf{u}_{\mathrm{A}, n}\Vert}; \\
		h_{\mathrm{R_{1}}\mathrm{R_{2}}, q_{2} q_{1}}  = \frac{\lambda \beta_{\mathrm{R_{1}}\mathrm{R_{2}}} }{4 \pi \Vert \mathbf{u}_{\mathrm{R}_{2}, q_{2}}-\mathbf{u}_{\mathrm{R}_{1}, q_{1}}\Vert} e^{-j \frac{2 \pi }{\lambda} \Vert \mathbf{u}_{\mathrm{R}_{2}, q_{2}}-\mathbf{u}_{\mathrm{R}_{1}, q_{1}}\Vert}; \\
		h_{\mathrm{R_{2}B}, m q_{2}}  = \frac{\lambda \beta_{\mathrm{R_{2}B}} }{4 \pi \Vert \mathbf{u}_{\mathrm{B}, m}-\mathbf{u}_{\mathrm{R_{2}}, q_{2}}\Vert} e^{-j \frac{2 \pi }{\lambda}\Vert \mathbf{u}_{\mathrm{B}, m}-\mathbf{u}_{\mathrm{R}, q_{2}}\Vert}; \\
		h_{\mathrm{AE}, v n} = \frac{\lambda \beta_{\mathrm{AE}} }{4 \pi \Vert \mathbf{u}_{\mathrm{E}, v}- \mathbf{u}_{\mathrm{A}, n}\Vert} e^{-j \frac{2 \pi }{\lambda}\Vert \mathbf{u}_{\mathrm{E}, v}- \mathbf{u}_{\mathrm{A}, n}\Vert}; \\
		h_{\mathrm{R_{1}E}, v q_{1}} = \frac{\lambda \beta_{\mathrm{R_{1}E}} }{4 \pi \Vert \mathbf{u}_{\mathrm{E}, v}-\mathbf{u}_{\mathrm{R_{1}}, q_{1}}\Vert} e^{-j \frac{2 \pi }{\lambda}\Vert \mathbf{u}_{\mathrm{E}, v}-\mathbf{u}_{\mathrm{R_{1}}, q_{1}}\Vert},
	\end{array}
	\right.
\label{eq:channel}
\end{equation}
where $\beta_{\mathrm{AR_{1}}}$, $\beta_{\mathrm{R_{1}}\mathrm{R_{2}}}$, $\beta_{\mathrm{R_{2}B}}$, $\beta_{\mathrm{AE}}$, and $\beta_{\mathrm{R_{2}E}}$ represent the attenuation from Alice to RIS$_{1}$, RIS$_{1}$ to RIS$_{2}$, RIS$_{2}$ to Bob, Alice to Eve, and RIS$_{2}$ to Eve, respectively.

We denote by $ \mathbf{x}_{\mathrm{R}_{2}}=\mathbf{H}_{\mathrm{R_{1}}\mathrm{R_{2}}}\boldsymbol{\Theta}_{1} \mathbf{H}_{\mathrm{AR_{1}}} \mathbf{x}$ as the incident signal on RIS$_{2}$. Then the reflected signal by RIS$_{2}$, denoted by $\mathbf{y}_{\mathrm{R}_{2}} $,  can be calculated as follows:
	\begin{equation}
		\mathbf{y}_{\mathrm{R}_{2}} = \mathbf{\Theta}_{2}\mathbf{A}\left(\mathbf{x}_{\mathrm{R}_{2}}+\mathbf{n}_{\mathrm{R}_{2}} \right),
	\end{equation}
where $\mathbf{n}_{\mathrm{R}_{2}} \sim \mathcal{C N}\left(0, \sigma_{\mathrm{R}_{2}}^{2} \mathbf{I}_{\mathrm{R}_{2}}\right)$ is the introduced noise at RIS$_{2}$ with the variance $ \sigma_{\mathrm{R}_{2}}^{2}$. Also, the total transmit power of RIS$_{2}$ is limited by its power supply given by
	\begin{equation}
		\begin{aligned}
		\mathbb{E}\left\lbrace \Vert  \mathbf{y}_{\mathrm{R}_{2}}\Vert^{2}\right\rbrace =&	\mathbb{E}\left\lbrace\Vert\mathbf{A}\left(\mathbf{x}_{\mathrm{R}_{2}}+\mathbf{n}_{\mathrm{R}_{2}} \right) \Vert^{2}\right\rbrace \\ =&\sum\limits_{l^\mathrm{s}_{\bar{n}}\in \mathcal{L}_\mathrm{s}}\hspace{-1mm} p_{l^\mathrm{s}_{\bar{n}}} \Vert\mathbf{A}\mathbf{H}_{\mathrm{R}_{2}}\mathbf{f}_{l^\mathrm{s}_{\bar{n}}}\Vert^{2} \hspace{-0.5mm}+\hspace{-0.5mm} \sigma_\mathrm{\ddot{z}}^2\sum\limits_{l^\mathrm{z}_{\ddot{n}}\in \mathcal{L}_\mathrm{\ddot {z}}}\hspace{-1mm}  \Vert\mathbf{A}\mathbf{H}_{\mathrm{R}_{2}}\mathbf{f}_{l^\mathrm{z}_{\ddot{n}}}\Vert^{2}\nonumber \\
		&+ \sigma_{\mathrm{R}_{2}}^{2}\Vert\mathbf{A}\Vert^{2}  \leq P_{\mathrm{R}_{2}},
		\end{aligned}
	\end{equation}
where $P_{\mathrm{R}_{2}} $ denotes the total effective transmit radio power of RIS$_{2}$ and $\mathbf{H}_{\mathrm{R}_{2}}= \mathbf{H}_{\mathrm{R_{1}}\mathrm{R_{2}}}\boldsymbol{\Theta}_{1} \mathbf{H}_{\mathrm{AR_{1}}}$.

Based on Eqs.~\eqref{eq:transmit signal} and~\eqref{eq:channel}, the received signals, denoted $\mathbf{y}_{\mathrm{B}}$ and $\mathbf{y}_{\mathrm{E}}$, for the whole OAM modes at Bob and Eve can be respectively obtained by
\begin{equation}
\mathbf{y}_{\mathrm{B}} = \mathbf{H}_{\mathrm{R_{2}B}} \boldsymbol{\Theta}_{2}\mathbf{A} \mathbf{H}_{\mathrm{R_{1}}\mathrm{R_{2}}}\boldsymbol{\Theta}_{1} \mathbf{H}_{\mathrm{AR_{1}}} \mathbf{x}+\mathbf{H}_{\mathrm{R_{2}B}} \boldsymbol{\Theta}_{2}\mathbf{A}\mathbf{n}_{\mathrm{R}_{2}}+\mathbf{n}_{\mathrm{B}}
\label{eq:y_b}
\end{equation}
and
\begin{equation}
\mathbf{y}_{\mathrm{E}} = \left(\mathbf{H}_{\mathrm{AE}}+\mathbf{H}_{\mathrm{R_{1}E}} \boldsymbol{\Theta}_{1} \mathbf{H}_{\mathrm{AR_{1}}}\right) \mathbf{x}+\mathbf{n}_{\mathrm{E},
\label{eq:y_E}}
\end{equation}
where  $\mathbf{n}_\mathrm{B} \sim \mathcal{C N}\left(0, \sigma_\mathrm{B}^{2} \mathbf{I}_{N}\right)$ and $\mathbf{n}_\mathrm{E} \sim \mathcal{C N}\left(0, \sigma_\mathrm{E}^{2} \mathbf{I}_{N_{\rm E}}\right)$ are the received noise at Bob with the variance $\sigma_\mathrm{B}^{2}$ and at Eve with the variance $\sigma_\mathrm{E}^{2}$, respectively.

To decompose OAM signals, the DFT operation is required at Bob. Thus, based on Eqs.~\eqref{eq:sz} and~\eqref{eq:y_b}, we can derive the decomposed signals, denoted by $\tilde{\mathbf{y}}_{\mathrm{B}}$, corresponding to $\mathbf{y}_{\mathrm{B}}$ at Bob as follows:
\begin{eqnarray}
	\tilde{\mathbf{y}}_{\mathrm{B}}=\hspace{-0.2cm}&\mathbf{F}^\mathrm{H}\mathbf{H}_{\mathrm{R_{2}B}} \boldsymbol{\Theta}_{2}\mathbf{A} \mathbf{H}_{\mathrm{R_{1}}\mathrm{R_{2}}}\boldsymbol{\Theta}_{1} \mathbf{H}_{\mathrm{AR_{1}}}  \mathbf{F}\left( \mathbf{s}+\mathbf{z}\right)
\nonumber    \\
&+\mathbf{F}^\mathrm{H}\mathbf{H}_{\mathrm{R_{2}B}} \boldsymbol{\Theta}_{2}\mathbf{A}\mathbf{n}_{\mathrm{R}_{2}}+\tilde{\mathbf{n}}_{\mathrm{B}}, \label{eq:DFT-yB}
\end{eqnarray}
where $\tilde{\mathbf{n}}_{\mathrm{B}}=\mathbf{F}^\mathrm{H} \mathbf{n}_{\mathrm{B}}$ follows the Gaussian distribution with variance $ \sigma_{\mathrm{B}}^{2}$.

Supposing that Alice and Bob can be completely synchronized, thus the corresponding SN pairs at Bob can be obtained based on the pre-shared index information. On the one hand, according to the obtained OAM index information for AN transmission and the pre-known AN by Alice and Bob, AN can be easily eliminated in Eq.~\eqref{eq:DFT-yB}. On the other hand, Bob can synchronously select the OAM mode set $\mathcal{L}_\mathrm{s}$ corresponding to desired signal transmissions according to the index information. For the OAM modes in $\mathcal{L}_\mathrm{s}$, RIS$_{1}$ and RIS$_{2}$ can jointly mitigate inter-mode interference caused by the misalignment between Alice and Bob. However, there still exists residual interference after DFT processing in the OAM wireless communications. Thereby, the signal-to-interference-plus-noise ratio (SINR), denoted by $\gamma_{\mathrm{B}, l^\mathrm{s}_{\bar{n}}}$, with respect to the OAM mode $l^\mathrm{s}_{\bar{n}}$ for Bob can be derived as follows:
\begin{equation}
		\gamma_{\mathrm{B}, l^\mathrm{s}_{\bar{n}}} = \frac{p_{l^\mathrm{s}_{\bar{n}}}\left| h^{\mathrm{B}}_{l^\mathrm{s}_{\bar{n}},l^\mathrm{s}_{\bar{n}}} \right|^{2}}{\sum\limits_{ k \neq l^\mathrm{s}_{\bar{n}}, k\in \mathcal{L}_\mathrm{s}} \!\!\!\!p_{k}\left| h^{\mathrm{B}}_{l^\mathrm{s}_{\bar{n}},k} \right|^{2}+\sigma_{\mathrm{R}_{2}}^{2}\Vert\mathbf{h}_{\mathrm{B},l^\mathrm{s}_{\bar{n}}}\mathbf{A}\Vert^{2}+\sigma_{\mathrm{B}}^{2}},
	\label{eq:gamma_B}
\end{equation}
where $h^{\mathrm{B}}_{l^\mathrm{s}_{\bar{n}},k}= \mathbf{f}_{l^\mathrm{s}_{\bar{n}}}^\mathrm{H}\mathbf{H}_{\mathrm{R_{2}B}} \boldsymbol{\Theta}_{2}\mathbf{A} \mathbf{H}_{\mathrm{R_{1}}\mathrm{R_{2}}}\boldsymbol{\Theta}_{1} \mathbf{H}_{\mathrm{AR_{1}}}  \mathbf{f}_{k}$ with $k\in \mathcal{L}_\mathrm{s}$ and $\mathbf{h}_{\mathrm{B},l^\mathrm{s}_{\bar{n}}}=\mathbf{f}_{l^\mathrm{s}_{\bar{n}}}^\mathrm{H}\mathbf{H}_{\mathrm{R_{2}B}} \boldsymbol{\Theta}_{2}$.

To obtain the achievable rate of our proposed double-RIS-assisted OAM with JiMa scheme in the near field, mutual information is utilized. The mutual information, denoted by $\mathcal{I}(\tilde{\mathbf{s}}, \tilde{\mathbf{y}}_{\mathrm{B}})$, between the transmit signals and the decomposed signals at Bob is given by
\begin{eqnarray}
    \mathcal{I}(\tilde{\mathbf{s}}, \tilde{\mathbf{y}}_{\mathrm{B}})= \mathcal{I}(\bm{s},\tilde{\mathbf{y}}_{\mathrm{B}}| \mathcal{L}_\mathrm{s},\mathcal{L_\mathrm{\ddot{z}}})+\mathcal{I}( \mathcal{L}_\mathrm{s},\mathcal{L_\mathrm{\ddot{z}}},\tilde{\mathbf{y}}_{\mathrm{B}}),
    \label{eq:I_sy}
\end{eqnarray}
where $\mathcal{I}(\bm{s},\tilde{\mathbf{y}}_{\mathrm{B}}| \mathcal{L}_\mathrm{s},\mathcal{L_\mathrm{\ddot{z}}})$ is the signal information and $\mathcal{I}(\mathcal{L}_\mathrm{s},\mathcal{L_\mathrm{\ddot{z}}},\tilde{\mathbf{y}}_{\mathrm{B}})$ is the index information. In the following, we have the achievable rate at Bob as Theorem 1.

{\textit{Theorem 1}:} The upper bound of achievable rate, denoted by $C_{\rm B}$, of our proposed double-RIS-assisted OAM with JiMa scheme at Bob is calculated by
\begin{equation}
C_{\mathrm{B}}=\sum\limits_{l^\mathrm{s}_{\bar{n}}\in \mathcal{L}_\mathrm{s}} \log _{2}\left(1+\gamma_{\mathrm{B}, l^\mathrm{s}_{\bar{n}}}\right)+\log_{2}G,
\label{eq:CB}
\end{equation}
where $\log_{2}G$ is the upper bound of $\mathcal{I}(\mathcal{L}_\mathrm{s},\mathcal{L_\mathrm{\ddot{z}}},\tilde{\mathbf{y}}_{\mathrm{B}})$.
\begin{proof}
    See Appendix~\ref{app:T1}.
\end{proof}

Different from Bob, Eve has no information about the emitted signals by Alice. Thus, Eve cannot distinguish the desired signals and AN in ${\mathbf{y}_{\rm E}}$. Also, Eve mainly focuses on working with the conventional MIMO. Hence, based on Eq.~\eqref{eq:y_E}, we can derive the SINR, denoted by $\gamma_{\mathrm{E}, l^\mathrm{s}_{\bar{n}}}$, corresponding to the OAM mode $l^\mathrm{s}_{\bar{n}}$ for Eve as follows:
\begin{equation}
		\gamma_{\mathrm{E}, l^\mathrm{s}_{\bar{n}}} = \frac{p_{l^\mathrm{s}_{\bar{n}}}\left| h^{\mathrm{E}}_{l^\mathrm{s}_{\bar{n}},l^\mathrm{s}_{\bar{n}}} \right|^{2}}{\sum\limits_{k \neq l^\mathrm{s}_{\bar{n}},k\in \mathcal{L}_\mathrm{s}}\!\!\!\! p_{k}\left| h^{\mathrm{E}}_{l^\mathrm{s}_{\bar{n}},k} \right|^{2}+\sigma_\mathrm{\ddot{z}}^2\sum\limits_{l^\mathrm{z}_{\ddot{n}} \in \mathcal{L}_\mathrm{\ddot{z}}}\!\!\!\left|h^{\mathrm{E}}_{l^\mathrm{s}_{\bar{n}},l^\mathrm{z}_{\ddot{n}}} \right|^{2}+\sigma_{\mathrm{E}}^{2}},
\label{eq:gamma_E}
\end{equation}
where $h^{\mathrm{E}}_{l^\mathrm{s}_{\bar{n}},\ddot{k}}= \left(\mathbf{h}_{\mathrm{AE},l^\mathrm{s}_{\bar{n}}}+\mathbf{h}_{\mathrm{R_{1}E},l^\mathrm{s}_{\bar{n}}} \boldsymbol{\Theta}_{1} \mathbf{H}_{\mathrm{AR_{1}}}\right) \mathbf{f}_{\ddot{k}}$ with $\ddot{k}\in \left\lbrace k,l^\mathrm{z}_{\ddot{n}}\right\rbrace $, $\mathbf{h}_{\mathrm{AE},l^\mathrm{s}_{\bar{n}}}$ and $\mathbf{h}_{\mathrm{R_{1}E},l^\mathrm{s}_{\bar{n}}}$ respectively represent the $l^\mathrm{s}_{\bar{n}}$-th row of $\mathbf{H}_{\mathrm{AE}}$ and $\mathbf{H}_{\mathrm{R_{1}E}}$. Then, the achievable rate of Eve, denoted by $R_{\mathrm{E}}$, can be respectively calculated by
\begin{equation}
R_{\mathrm{E}}=\sum\limits_{l^\mathrm{s}_{\bar{n}}\in \mathcal{L}_\mathrm{s}} \log _{2}\left(1+\gamma_{\mathrm{E}, l^\mathrm{s}_{\bar{n}}}\right).
\label{eq:CE}
\end{equation}

\setcounter{equation}{25}
\begin{figure*}[hbp]
\hrulefill
  \begin{equation}
\varphi_{\mathrm{B},l^\mathrm{s}_{\bar{n}}}(\mathbf{p}_\mathrm{s},t_{\mathrm{B},l^\mathrm{s}_{\bar{n}}}) =\ln\!\left(\!1\!+\!\left( \sum\limits_{ k\in \mathcal{L}_\mathrm{s}} \!\! p_{k}\left| h^{\mathrm{B}}_{l^\mathrm{s}_{\bar{n}},k} \right|^{2}+c_{1,l^\mathrm{s}_{\bar{n}}}\right)\sigma_{\mathrm{B}}^{-2}\!\right) -t_{\mathrm{B},l^\mathrm{s}_{\bar{n}}}\!\left(\hspace{-0.5mm}\!1\!+\left( \sum\limits_{ k \neq l^\mathrm{s}_{\bar{n}}, k\in \mathcal{L}_\mathrm{s}} \hspace{-5mm} p_{k}\left| h^{\mathrm{B}}_{l^\mathrm{s}_{\bar{n}},k} \right|^{2}\hspace{-1mm}+\hspace{-0.5mm}c_{1,l^\mathrm{s}_{\bar{n}}}\hspace{-1mm}\right)\sigma_{\mathrm{B}}^{-2}\hspace{-0.5mm}\right)\hspace{-0.5mm}+\hspace{-0.5mm}\ln t_{\mathrm{B},l^\mathrm{s}_{\bar{n}}}\hspace{-0.5mm}+\hspace{-0.5mm}1.
\label{eq:phi_B}
\end{equation}
\end{figure*}

\setcounter{equation}{27}
\begin{figure*} [hbp]
\hrulefill
  \begin{equation}
\varphi_{\mathrm{E},l^\mathrm{s}_{\bar{n}}}(\mathbf{p}_\mathrm{s},t_{\mathrm{E},l^\mathrm{s}_{\bar{n}}})\hspace{-0.5mm}=\hspace{-0.5mm} t_{\mathrm{E},l^\mathrm{s}_{\bar{n}}}\!\left(\hspace{-0.5mm}1+\left( \sum\limits_{ k\in \mathcal{L}_\mathrm{s}} \!\! p_{k}\left| h^{\mathrm{E}}_{l^\mathrm{s}_{\bar{n}},k} \right|^{2}+c_{2,l^\mathrm{s}_{\bar{n}}}\right)\sigma_{\mathrm{E}}^{-2}\hspace{-0.5mm}\right)
-
\ln\!\left(\hspace{-0.5mm}1\!+\left( \sum\limits_{ k \neq l^\mathrm{s}_{\bar{n}}, k\in \mathcal{L}_\mathrm{s}} \hspace{-4mm} p_{k}\left| h^{\mathrm{E}}_{l^\mathrm{s}_{\bar{n}},k} \right|^{2}+c_{2,l^\mathrm{s}_{\bar{n}}}\right)\sigma_{\mathrm{E}}^{-2}\hspace{-0.5mm}\right)
\hspace{-0.5mm}-\hspace{-0.5mm}\ln t_{\mathrm{E},l^\mathrm{s}_{\bar{n}}}\hspace{-0.5mm}-\hspace{-0.5mm}1.
\label{eq:phi_E}
\end{equation}
\end{figure*}

\subsection{Secrecy Rate Optimization Problem Formulation}
To evaluate the secure performance in the near field, the secrecy rate, denoted by $R_{\rm OAM}$, of our proposed scheme can be calculated by
\setcounter{equation}{18}
\begin{equation}
R_{\rm OAM} = C_{\mathrm{B}}-R_{\mathrm{E}}.
\end{equation}
Aiming at maximizing the secrecy rate, we jointly optimize the transmit power allocation $\mathbf{p}_\mathrm{s}=[p_{l^\mathrm{s}_{1}},\ldots,p_{l^\mathrm{s}_{N_\mathrm{s}}} ]^{\mathrm{T}}$ for desired signal transmissions, the phase shift $\boldsymbol{\theta}_{1}$ of RIS$_{1}$, and phase shift $\boldsymbol{\theta}_{2}$ of RIS$_{2}$ subject to the constraints of total transmit power of desired signals and unit modulus of phase shifts. Therefore, the secrecy rate optimization problem can be formulated as follows:

\begin{subequations}
\begin{align}	
		\textbf{P1}: \; \max _{\mathbf{p}_\mathrm{s},\mathbf{a}, \boldsymbol{\theta}_{1},\boldsymbol{\theta}_{2}} \;&   R_{\mathrm{OAM}}\\
		\mathrm{s.t}. \quad  & \sum\limits_{l^\mathrm{s}_{\bar{n}}\in \mathcal{L}_\mathrm{s}} p_{l^\mathrm{s}_{\bar{n}}} \leq \rho P_\mathrm{T} ;\\
		&p_{l^\mathrm{s}_{\bar{n}}} \geq p_\mathrm{th}, \forall l^\mathrm{s}_{\bar{n}} \in \mathcal{L}_\mathrm{s};\\
		& \sum\limits_{l^\mathrm{s}_{\bar{n}}\in \mathcal{L}_\mathrm{s}}\hspace{-1mm} p_{l^\mathrm{s}_{\bar{n}}} \Vert\mathbf{A}\mathbf{H}_{\mathrm{R}_{2}}\mathbf{f}_{l^\mathrm{s}_{\bar{n}}}\Vert^{2} \hspace{-0.5mm}+\hspace{-0.5mm} \sigma_\mathrm{\ddot{z}}^2\sum\limits_{l^\mathrm{z}_{\ddot{n}}\in \mathcal{L}_\mathrm{\ddot {z}}}\hspace{-1mm}  \Vert\mathbf{A}\mathbf{H}_{\mathrm{R}_{2}}\mathbf{f}_{l^\mathrm{z}_{\ddot{n}}}\Vert^{2}\nonumber \\
		&+ \sigma_{\mathrm{R}_{2}}^{2}\Vert\mathbf{A}\Vert^{2} \leq P_{\mathrm{R}_{2}};\\
		& 0\leq a_{q_{2}}\leq a_{\mathrm{max}};\\
		&\left|\theta_{1,q_{1}}\right|  = 1, \forall q_{1} \in \left\lbrace 1,\dots, Q_{1}\right\rbrace;\\
		&\left|\theta_{2,q_{2}}\right|  = 1, \forall q_{2} \in \left\lbrace 1,\dots, Q_{2}\right\rbrace,
\end{align}	
\end{subequations}
where $ p_\mathrm{th}$ $\left( p_\mathrm{th}>0\right)$ is the allocated power threshold. Due to the non-convexity of problem $\textbf{P1}$ with coupled optimization variables and the non-convex unit modulus constraints, it is not easy to obtain the optimal solution. Therefore, it has an obligation to find an effective optimization algorithm to achieve the maximum secrecy rate of our proposed double-RIS-assisted OAM with JiMa scheme in the near field for anti-eavesdropping.

\section{Achieving the Maximum Secrecy Rate \\ with The RMCG-AO Algorithm}\label{sec:Algorithm}

In this section, the RMCG-AO algorithm is developed to tackle problem $\textbf{P1}$. Specifically, we first decompose $\textbf{P1}$ into four subproblems: 1) optimize $\mathbf{p}_\mathrm{s}$ with given $\mathbf{a}$, $ \boldsymbol{\theta}_{1} $, and $ \boldsymbol{\theta}_{2} $; 2) optimize $\mathbf{a}$ with given $\mathbf{p}_\mathrm{s}$, $ \boldsymbol{\theta}_{1} $, and $ \boldsymbol{\theta}_{2} $; 3) optimize $ \boldsymbol{\theta}_{1} $ with given $ \mathbf{p}_\mathrm{s} $, $\mathbf{a}$, and $ \boldsymbol{\theta}_{2} $; 4) optimize $ \boldsymbol{\theta}_{2} $ with given $ \mathbf{p}_\mathrm{s} $, $\mathbf{a}$, and $ \boldsymbol{\theta}_{1} $. Then, we alternately solve these four  subproblems.

\subsection{Optimize $\mathbf{p}_\mathrm{s}$ with given $\mathbf{a}$, $ \boldsymbol{\theta}_{1} $, and $ \boldsymbol{\theta}_{2} $}
For the convenience of analysis, we have $R_{\mathrm{B}}=C_{\mathrm{B}}-\log_{2} G$. For given $\mathbf{a}$, $\boldsymbol{\theta}_{1}$, and $ \boldsymbol{\theta}_{2} $,  $R_{\mathrm{B}}$ can be re-expressed as follows:
\begin{equation}
	\begin{aligned}
R_{\mathrm{B}}=& \sum_{l^\mathrm{s}_{\bar{n}}\in \mathcal{L}_\mathrm{s}} \left[ \log _{2}\left(1+\left( \sum\limits_{ k\in \mathcal{L}_\mathrm{s}} \!\! p_{k}\left| h^{\mathrm{B}}_{l^\mathrm{s}_{\bar{n}},k} \right|^{2}+c_{1,l^\mathrm{s}_{\bar{n}}}\right)\sigma_{\mathrm{B}}^{-2}\right) \right.\\
&\left.-\log _{2}\left(1+\left( \sum\limits_{ k \neq l^\mathrm{s}_{\bar{n}}, k\in \mathcal{L}_\mathrm{s}} \hspace{-4mm} p_{k}\left| h^{\mathrm{B}}_{l^\mathrm{s}_{\bar{n}},k} \right|^{2}+c_{1,l^\mathrm{s}_{\bar{n}}}\right)\sigma_{\mathrm{B}}^{-2} \right)\right],
\end{aligned}
\end{equation}
where $c_{1,l^\mathrm{s}_{\bar{n}}}= \sigma_{\mathrm{R}_{2}}^{2}\Vert\mathbf{h}_{\mathrm{B},l^\mathrm{s}_{\bar{n}}}\mathbf{A}\Vert^{2}$. Similarly,  $R_{\mathrm{E}}$ in Eq.~\eqref{eq:CE} can be re-written by
\begin{equation}
	\begin{aligned}
		R_{\mathrm{E}}=& \sum_{l^\mathrm{s}_{\bar{n}}\in \mathcal{L}_\mathrm{s}} \left[ \log _{2}\left(1+\left( \sum\limits_{ k\in \mathcal{L}_\mathrm{s}} \!\! p_{k}\left| h^{\mathrm{E}}_{l^\mathrm{s}_{\bar{n}},k} \right|^{2}+c_{2,l^\mathrm{s}_{\bar{n}}}\right)\sigma_{\mathrm{E}}^{-2}\right) \right.\\
		&\left.-\log _{2}\left(1+\left( \sum\limits_{ k \neq l^\mathrm{s}_{\bar{n}}, k\in \mathcal{L}_\mathrm{s}} \hspace{-4mm} p_{k}\left| h^{\mathrm{E}}_{l^\mathrm{s}_{\bar{n}},k} \right|^{2}+c_{2,l^\mathrm{s}_{\bar{n}}}\right)\sigma_{\mathrm{E}}^{-2} \right)\right],
	\end{aligned}
\end{equation}
where $c_{2,l^\mathrm{s}_{\bar{n}}}=\sigma_\mathrm{\ddot{z}}^2\sum\limits_{l^\mathrm{z}_{\ddot{n}} \in \mathcal{L}_\mathrm{\ddot{z}}}\!\!\!\left|h^{\mathrm{E}}_{l^\mathrm{s}_{\bar{n}},l^\mathrm{z}_{\ddot{n}}} \right|^{2}$.
Consequently, problem $ \textbf{P1} $ can be re-formulated as follows:
\begin{subequations}
		\begin{align}	
		\textbf{P2}: \;\  \max _{\mathbf{p}_\mathrm{s} }\; &R_{\mathrm{B}}-R_{\mathrm{E}} \\
		\mathrm{s.t}.\;\, & \sum\limits_{l^\mathrm{s}_{\bar{n}}\in \mathcal{L}_\mathrm{s}} p_{l^\mathrm{s}_{\bar{n}}} \leq \rho P_\mathrm{T}; \\  &p_{l^\mathrm{s}_{\bar{n}}} \geq p_\mathrm{th}, \forall l^\mathrm{s}_{\bar{n}}\in \mathcal{L}_\mathrm{s};\\
		& \sum\limits_{l^\mathrm{s}_{\bar{n}}\in \mathcal{L}_\mathrm{s}}\hspace{-1mm} p_{l^\mathrm{s}_{\bar{n}}} \Vert\mathbf{A}\mathbf{H}_{\mathrm{R}_{2}}\mathbf{f}_{l^\mathrm{s}_{\bar{n}}}\Vert^{2} \hspace{-0.5mm}+\hspace{-0.5mm} \sigma_\mathrm{\ddot{z}}^2\sum\limits_{l^\mathrm{z}_{\ddot{n}}\in \mathcal{L}_\mathrm{\ddot {z}}}\hspace{-1mm}  \Vert\mathbf{A}\mathbf{H}_{\mathrm{R}_{2}}\mathbf{f}_{l^\mathrm{z}_{\ddot{n}}}\Vert^{2}\nonumber \\
		&+ \sigma_{\mathrm{R}_{2}}^{2}\Vert\mathbf{A}\Vert^{2} \leq P_{\mathrm{R}_{2}}.
	\end{align}	
\end{subequations}
Clearly, $ \textbf{P2} $ is a non-convex problem. \textbf{Lemma 1} is applied to solve the non-convexity~\cite{Transmit_Solutions_for_MIMO_AO}.

\begin{figure*}[hbp]
	\setcounter{equation}{32}
	\hrulefill
	\begin{equation}
		\begin{aligned}
			&\log_{2}\left(1+\gamma_{\mathrm{B},l^\mathrm{s}_{\bar{n}}}\right)\\
&=\frac{1}{\ln 2} \max_{\omega_{l^\mathrm{s}_{\bar{n}}}>0,\tau_{l^\mathrm{s}_{\bar{n}}}}
			\ln\left(\omega_{l^\mathrm{s}_{\bar{n}}} \right) -\omega_{l^\mathrm{s}_{\bar{n}}}e_{l^\mathrm{s}_{\bar{n}}}+1 \\
			&=\frac{1}{\ln 2} \max_{\omega_{l^\mathrm{s}_{\bar{n}}}>0,\tau_{l^\mathrm{s}_{\bar{n}}}}\hspace{-2mm}\ln   \left(\omega_{l^\mathrm{s}_{\bar{n}}} \right)\hspace{-0.5mm}-\hspace{-0.5mm}\omega_{l^\mathrm{s}_{\bar{n}}}\hspace{-1mm}	\left[\hspace{-1mm}\left(\hspace{-1mm} 1\hspace{-0.5mm}-\hspace{-0.5mm}\tau_{l^\mathrm{s}_{\bar{n}}}^{*} \sqrt{p_{l^\mathrm{s}_{\bar{n}}}} h^{\mathrm{B}}_{l^\mathrm{s}_{\bar{n}},l^\mathrm{s}_{\bar{n}}}\hspace{-0.5mm}\right) \hspace{-0.5mm}\left(\hspace{-1mm} 1\hspace{-0.5mm}-\hspace{-0.5mm}\tau_{l^\mathrm{s}_{\bar{n}}}^{*} \sqrt{p_{l^\mathrm{s}_{\bar{n}}}} h^{\mathrm{B}}_{l^\mathrm{s}_{\bar{n}},l^\mathrm{s}_{\bar{n}}}\hspace{-0.5mm}\right)^{\mathrm{H}}
			\hspace{-1mm}+\hspace{-0.5mm}\left| \tau_{l^\mathrm{s}_{\bar{n}}}\right|^{2}\hspace{-1mm}\left( \sum\limits_{ k \neq l^\mathrm{s}_{\bar{n}}, k\in \mathcal{L}_\mathrm{s}} \hspace{-4mm} p_{k}\left| h^{\mathrm{B}}_{l^\mathrm{s}_{\bar{n}},k} \right|^{2}\hspace{-1mm}\hspace{-0.5mm}+\hspace{-0.5mm}\hspace{-0.5mm}\sigma_{\mathrm{R}_{2}}^{2}\Vert\mathbf{h}_{\mathrm{B},l^\mathrm{s}_{\bar{n}}}\mathbf{A}\Vert^{2}\hspace{-0.5mm}+\hspace{-0.5mm}\sigma_{\mathrm{B}}^{2}\hspace{-0.5mm}\right)\hspace{-1mm} \right] \hspace{-1.5mm}+\hspace{-1mm}1	\\
			&=\frac{1}{\ln 2} \max_{\omega_{l^\mathrm{s}_{\bar{n}}}>0,\tau_{l^\mathrm{s}_{\bar{n}}}}\hspace{-2mm}\underbrace{\ln\left(\hspace{-0.5mm}\omega_{l^\mathrm{s}_{\bar{n}}}\hspace{-0.5mm} \right)\hspace{-0.5mm}-\hspace{-0.5mm}\omega_{l^\mathrm{s}_{\bar{n}}}\hspace{-0.5mm}-\hspace{-0.5mm}\omega_{l^\mathrm{s}_{\bar{n}}}\hspace{-0.5mm}\left| \tau_{l^\mathrm{s}_{\bar{n}}}\right|^{2}\hspace{-0.5mm}\sum\limits_{ k\in \mathcal{L}_\mathrm{s}} \hspace{-1mm} p_{k}\hspace{-0.5mm}\left| h^{\mathrm{B}}_{l^\mathrm{s}_{\bar{n}},k} \right|^{2}\hspace{-0.5mm}+\hspace{-0.5mm}2\omega_{l^\mathrm{s}_{\bar{n}}}\hspace{-0.5mm}\Re\left\lbrace\hspace{-0.5mm} \tau_{l^\mathrm{s}_{\bar{n}}}^{*} \sqrt{p_{l^\mathrm{s}_{\bar{n}}}} h^{\mathrm{B}}_{l^\mathrm{s}_{\bar{n}},l^\mathrm{s}_{\bar{n}}} \hspace{-0.5mm}\right\rbrace \hspace{-0.5mm}-\hspace{-0.5mm} \omega_{l^\mathrm{s}_{\bar{n}}}\hspace{-0.5mm}\left| \tau_{l^\mathrm{s}_{\bar{n}}}\right|^{2}\hspace{-1mm}\sigma_{\mathrm{R}_{2}}^{2}\Vert\mathbf{h}_{\mathrm{B},l^\mathrm{s}_{\bar{n}}}\mathbf{A}\Vert^{2}\hspace{-0.5mm}-\hspace{-0.5mm}\omega_{l^\mathrm{s}_{\bar{n}}}\hspace{-0.5mm}\left| \tau_{l^\mathrm{s}_{\bar{n}}}\right|^{2}\hspace{-0.5mm}\sigma_{\mathrm{B}}^{2}\hspace{-0.5mm}+\hspace{-0.5mm}1}_{ f_{\mathrm{B},l^\mathrm{s}_{\bar{n}}}},
		\end{aligned}	
		\label{eq:f_B}
	\end{equation}
	
\end{figure*}

\textbf{Lemma 1}: For any constant $x>0$, we have the maximum of function $\Theta(t)=-t x+\ln t+1$ versus the auxiliary variable $t$ as follows:
\setcounter{equation}{23}
\begin{equation}
-\ln x=\max _{t>0} \Theta(t).
\label{lemma1}
\end{equation}
The optimal solution of $\Theta(t)$ is $1/x$.

We set $ x\hspace{-0.5mm}=\hspace{-0.5mm}1\hspace{-0.5mm}+\hspace{-0.5mm}\left( \sum\limits_{ k \neq l^\mathrm{s}_{\bar{n}}, k\in \mathcal{L}_\mathrm{s}} \hspace{-5mm} p_{k}\left| h^{\mathrm{B}}_{l^\mathrm{s}_{\bar{n}},k} \right|^{2}\hspace{-1mm}+\hspace{-0.5mm}c_{1,l^\mathrm{s}_{\bar{n}}}\hspace{-1mm}\right)\sigma_{\mathrm{B}}^{-2} $
and $ t=t_{\mathrm{B},l^\mathrm{s}_{\bar{n}}} $ according to \textbf{Lemma 1}, thus calculating $  R_{\mathrm{B}}$ as follows:
\begin{equation}
R_{\mathrm{B}}= \frac{1}{\ln 2}\sum\limits_{l^\mathrm{s}_{\bar{n}}\in \mathcal{L}_\mathrm{s}} \max _{t_{\mathrm{B},l^\mathrm{s}_{\bar{n}}}>0} \varphi_{\mathrm{B},l^\mathrm{s}_{\bar{n}}}(\mathbf{p}_\mathrm{s},t_{\mathrm{B},l^\mathrm{s}_{\bar{n}}}),
\label{con:Rbln2}
\end{equation}
where $\varphi_{\mathrm{B},l^\mathrm{s}_{\bar{n}}}(\mathbf{p}_\mathrm{s},t_{\mathrm{B},l^\mathrm{s}_{\bar{n}}})$ is derived as shown in Eq.~\eqref{eq:phi_B}.
With the similar method, we can calculate $R_{\mathrm{E}}$ as follows:
\setcounter{equation}{26}
\begin{equation}
R_{\mathrm{E}}= \frac{1}{\ln 2} \sum\limits_{l^\mathrm{s}_{\bar{n}}\in \mathcal{L}_\mathrm{s}} \min _{t_{\mathrm{E},l^\mathrm{s}_{\bar{n}}}>0} \varphi_{\mathrm{E},l^\mathrm{s}_{\bar{n}}}(\mathbf{p}_\mathrm{s},t_{\mathrm{E},l^\mathrm{s}_{\bar{n}}}),
\label{con:Reln2}
\end{equation}
where $\varphi_{\mathrm{E},l^\mathrm{s}_{\bar{n}}}(\mathbf{p}_\mathrm{s},t_{\mathrm{E},l^\mathrm{s}_{\bar{n}}})$ is given as shown in Eq.~\eqref{eq:phi_E}.

Hence, we can transform problem $ \textbf{P2} $ into $ \textbf{P3} $ as follows:
\setcounter{equation}{28}
\begin{subequations}
\begin{align}
\textbf{P3} : \!\max _{\mathbf{p}_\mathrm{s} ,t_{\mathrm{B},l^\mathrm{s}_{\bar{n}}}\!\!,\atop t_{\mathrm{E},l^\mathrm{s}_{\bar{n}}}} & \sum\limits_{l^\mathrm{s}_{\bar{n}}\in \mathcal{L}_\mathrm{s}}\hspace{-2mm}\varphi_{\mathrm{B},l^\mathrm{s}_{\bar{n}}}\hspace{-0.5mm}\left( \mathbf{p}_\mathrm{s},t_{\mathrm{B},l^\mathrm{s}_{\bar{n}}}\right) \hspace{-1mm}-\hspace{-1mm}\sum\limits_{l^\mathrm{s}_{\bar{n}}\in \mathcal{L}_\mathrm{s}}\hspace{-2mm}\varphi_{\mathrm{E},l^\mathrm{s}_{\bar{n}}}\!\!\left( \mathbf{p}_\mathrm{s},t_{\mathrm{E},l^\mathrm{s}_{\bar{n}}}\right)  \\
\mathrm{s.t}.\quad \;\;\, & \sum\limits_{l^\mathrm{s}_{\bar{n}}\in \mathcal{L}_\mathrm{s}} p_{l^\mathrm{s}_{\bar{n}}} \leq \rho P_\mathrm{T};\\ &p_{l^\mathrm{s}_{\bar{n}}} \geq p_\mathrm{th};\\
& \sum\limits_{l^\mathrm{s}_{\bar{n}}\in \mathcal{L}_\mathrm{s}}\hspace{-1mm} p_{l^\mathrm{s}_{\bar{n}}} \Vert\mathbf{A}\mathbf{H}_{\mathrm{R}_{2}}\mathbf{f}_{l^\mathrm{s}_{\bar{n}}}\Vert^{2} \hspace{-0.5mm}+\hspace{-0.5mm} \sigma_\mathrm{\ddot{z}}^2\sum\limits_{l^\mathrm{z}_{\ddot{n}}\in \mathcal{L}_\mathrm{\ddot {z}}}\hspace{-1mm}  \Vert\mathbf{A}\mathbf{H}_{\mathrm{R}_{2}}\mathbf{f}_{l^\mathrm{z}_{\ddot{n}}}\Vert^{2}\nonumber \\
&+ \sigma_{\mathrm{R}_{2}}^{2}\Vert\mathbf{A}\Vert^{2} \leq P_{\mathrm{R}_{2}};\\
&t_{\mathrm{B},l^\mathrm{s}_{\bar{n}}},t_{\mathrm{E},l^\mathrm{s}_{\bar{n}}}>0, \forall l^\mathrm{s}_{\bar{n}} \in \mathcal{L}_\mathrm{s}.
\end{align}
\end{subequations}
It can be demonstrated that $ \textbf{P3} $ is convex problem about $\mathbf{p}_\mathrm{s}$ and $\left(t_{\mathrm{B},l^\mathrm{s}_{\bar{n}}},t_{\mathrm{E},l^\mathrm{s}_{\bar{n}}}\right)$. The optimal solution can be obtained with the alternative optimization algorithm. Therefore, based on \textbf{Lemma 1}, the optimal solutions of $\left(t_{\mathrm{B},l^\mathrm{s}_{\bar{n}}},t_{\mathrm{E},l^\mathrm{s}_{\bar{n}}}\right)$, denoted by $\left(t_{\mathrm{B},l^\mathrm{s}_{\bar{n}}}^{\star},t_{\mathrm{E},l^\mathrm{s}_{\bar{n}}}^{\star}\right)$, for given $ \mathbf{p}_\mathrm{s} $ can be calculated by
\begin{equation}
	\left\{
	\begin{array}{lr}
t_{\mathrm{B},l^\mathrm{s}_{\bar{n}}}^{\star}=\left (1+\left( \sum\limits_{ k \neq l^\mathrm{s}_{\bar{n}}, k\in \mathcal{L}_\mathrm{s}} \hspace{-5mm} p_{k}\left| h^{\mathrm{B}}_{l^\mathrm{s}_{\bar{n}},k} \right|^{2}\hspace{-1mm}+\hspace{-0.5mm}c_{1,l^\mathrm{s}_{\bar{n}}}\hspace{-1mm}\right)\sigma_{\mathrm{B}}^{-2}\right)^{-1};\\
	t_{\mathrm{E},l^\mathrm{s}_{\bar{n}}}^{\star}=\left (1+\left( \sum\limits_{ k\in \mathcal{L}_\mathrm{s}} \!\! p_{k}\left| h^{\mathrm{E}}_{l^\mathrm{s}_{\bar{n}},k} \right|^{2}+c_{2,l^\mathrm{s}_{\bar{n}}}\right)\sigma_{\mathrm{E}}^{-2}\right)^{-1}.
	\end{array}
	\right.
\label{eq:t_{B,E,l}}
\end{equation}
Then, we can obtain the optimal solution of $\mathbf{p}_\mathrm{s}$, denoted by $\mathbf{p}_\mathrm{s}^{\star}$, by solving problem $ \textbf{P4} $ given as follows:
\begin{subequations}
\begin{align}
\textbf{P4} : \; \max _{\mathbf{p}_\mathrm{s} }\hspace{-1mm} & \sum\limits_{l^\mathrm{s}_{\bar{n}}\in \mathcal{L}_\mathrm{s}}\hspace{-1mm}\varphi_{\mathrm{B},l^\mathrm{s}_{\bar{n}}}\left( \mathbf{p},t_{\mathrm{B},l^\mathrm{s}_{\bar{n}}}^{\star}\right)\hspace{-1mm} -\hspace{-2mm}\sum\limits_{l^\mathrm{s}_{\bar{n}}\in \mathcal{L}_\mathrm{s}}\varphi_{\mathrm{E},l^\mathrm{s}_{\bar{n}}}\left( \mathbf{p},t_{\mathrm{E},l^\mathrm{s}_{\bar{n}}}^{\star}\right)  \\
\mathrm{s.t}.\;  & \sum\limits_{l^\mathrm{s}_{\bar{n}}\in \mathcal{L}_\mathrm{s}} p_{l^\mathrm{s}_{\bar{n}}} \leq \rho P_\mathrm{T};\\
&  p_{l^\mathrm{s}_{\bar{n}}} \geq p_\mathrm{th}, \forall l^\mathrm{s}_{\bar{n}} \in \mathcal{L}_\mathrm{s};\\
& \sum\limits_{l^\mathrm{s}_{\bar{n}}\in \mathcal{L}_\mathrm{s}}\hspace{-1mm} p_{l^\mathrm{s}_{\bar{n}}} \Vert\mathbf{A}\mathbf{H}_{\mathrm{R}_{2}}\mathbf{f}_{l^\mathrm{s}_{\bar{n}}}\Vert^{2} \hspace{-0.5mm}+\hspace{-0.5mm} \sigma_\mathrm{\ddot{z}}^2\sum\limits_{l^\mathrm{z}_{\ddot{n}}\in \mathcal{L}_\mathrm{\ddot {z}}}\hspace{-1mm}  \Vert\mathbf{A}\mathbf{H}_{\mathrm{R}_{2}}\mathbf{f}_{l^\mathrm{z}_{\ddot{n}}}\Vert^{2}\nonumber \\
&+ \sigma_{\mathrm{R}_{2}}^{2}\Vert\mathbf{A}\Vert^{2} \leq P_{\mathrm{R}_{2}}.
\end{align}
\end{subequations}
$\textbf{P4}$ is a convex problem, thus easily being solved by convex problem solvers such as CVX.

Above all, problem $\textbf{P2} $ can be solved by alternately updating $ \left(t_{\mathrm{B},l^\mathrm{s}_{\bar{n}}},t_{\mathrm{E},l^\mathrm{s}_{\bar{n}}}\right)$ and $\mathbf{p}_\mathrm{s}$.

\subsection{Optimize $\mathbf{a}$ with given $\mathbf{p}_\mathrm{s}$, $ \boldsymbol{\theta}_{1} $, and $ \boldsymbol{\theta}_{2} $}
With given $\mathbf{p}_\mathrm{s}$, $ \boldsymbol{\theta}_{1} $, and $ \boldsymbol{\theta}_{2} $, we first introduce the auxiliary variables $\boldsymbol{\omega} =\left[\omega_{l^\mathrm{s}_{1}} ,\ldots,\omega_{l^\mathrm{s}_{N_\mathrm{s}}} \right]^{\mathrm{T}} $ and $\boldsymbol{\tau} =\left[\tau_{l^\mathrm{s}_{1}} ,\ldots,\tau_{l^\mathrm{s}_{N_\mathrm{s}}} \right]^{\mathrm{T}} $. The mean square error (MSE) function of Eq.~\eqref{eq:DFT-yB} can be defined as follows:
\begin{equation}
	\begin{aligned}
	&e_{l^\mathrm{s}_{\bar{n}}}\left( \mathbf{a},\tau_{l^\mathrm{s}_{\bar{n}}}\right) \\
	&=\left( 1-\tau_{l^\mathrm{s}_{\bar{n}}}^{*} \sqrt{p_{l^\mathrm{s}_{\bar{n}}}} h^{\mathrm{B}}_{l^\mathrm{s}_{\bar{n}},l^\mathrm{s}_{\bar{n}}}\right) \left( 1-\tau_{l^\mathrm{s}_{\bar{n}}}^{*} \sqrt{p_{l^\mathrm{s}_{\bar{n}}}} h^{\mathrm{B}}_{l^\mathrm{s}_{\bar{n}},l^\mathrm{s}_{\bar{n}}}\right)^{\mathrm{H}}  \\
	&+\left| \tau_{l^\mathrm{s}_{\bar{n}}}\right|^{2}\hspace{-1mm}\left( \sum\limits_{ k \neq l^\mathrm{s}_{\bar{n}}, k\in \mathcal{L}_\mathrm{s}} \hspace{-4mm} p_{k}\left| h^{\mathrm{B}}_{l^\mathrm{s}_{\bar{n}},k} \right|^{2}\hspace{-1mm}+\hspace{-0.5mm}\sigma_{\mathrm{R}_{2}}^{2}\Vert\mathbf{h}_{\mathrm{B},l^\mathrm{s}_{\bar{n}}}\mathbf{A}\Vert^{2}+\sigma_{\mathrm{B}}^{2}\right). \label{eq:MSE}
	\end{aligned}
\end{equation}
By using the Lemma 1 in \cite{10061167}, $\log _{2}\left(1+\gamma_{\mathrm{B}, l^\mathrm{s}_{\bar{n}}}\right) $ can be equivalently expressed as shown in Eq.~\eqref{eq:f_B}, where  $f_{\mathrm{B},l^\mathrm{s}_{\bar{n}}}$ is concave with respect to $\mathbf{a} $, $\omega_{l^\mathrm{s}_{\bar{n}}} $, and $\tau_{l^\mathrm{s}_{\bar{n}}} $ by fixing the other variables.

With $\mathbf{p}_\mathrm{s}$ and $\mathbf{a}$ being fixed, the MSE function in Eq.~\eqref{eq:MSE} is convex with respect to $\tau_{l^\mathrm{s}_{\bar{n}}}$. By checking the first-order optimality condition of $\tau_{l^\mathrm{s}_{\bar{n}}}$ in Eq.~\eqref{eq:MSE}, we can obtain the closed-form expression of $\tau_{l^\mathrm{s}_{\bar{n}}}$ as follows:
\setcounter{equation}{33}
\begin{equation}
	\tau_{l^\mathrm{s}_{\bar{n}}}=\frac{\sqrt{p_{l^\mathrm{s}_{\bar{n}}}}h^{\mathrm{B}}_{l^\mathrm{s}_{\bar{n}},l^\mathrm{s}_{\bar{n}}}}{\sum\limits_{ k\in \mathcal{L}_\mathrm{s}} \hspace{-1mm} p_{k}\left| h^{\mathrm{B}}_{l^\mathrm{s}_{\bar{n}},k} \right|^{2}\hspace{-1mm}+\hspace{-0.5mm}\sigma_{\mathrm{R}_{2}}^{2}\Vert\mathbf{h}_{\mathrm{B},l^\mathrm{s}_{\bar{n}}}\mathbf{A}\Vert^{2}+\sigma_{\mathrm{B}}^{2}}.\label{eq:tau}
\end{equation}

Then, according to the Lemma 1 in \cite{10061167}, with $\mathbf{a}$ and $\tau_{l^\mathrm{s}_{\bar{n}}}$ being fixed, the analytical
solution of $\omega_{l^\mathrm{s}_{\bar{n}}} $ can be expressed by
\begin{equation}
\begin{aligned}
	\omega_{l^\mathrm{s}_{\bar{n}}}&=\frac{1}{e_{l^\mathrm{s}_{\bar{n}}}\left( \mathbf{a},\tau_{l^\mathrm{s}_{\bar{n}}}\right)}\\
	&=1+\frac{p_{l^\mathrm{s}_{\bar{n}}}\left| h^{\mathrm{B}}_{l^\mathrm{s}_{\bar{n}},l^\mathrm{s}_{\bar{n}}} \right|^{2}}{\sum\limits_{ k \neq l^\mathrm{s}_{\bar{n}}, k\in \mathcal{L}_\mathrm{s}} \!\!\!\!p_{k}\left| h^{\mathrm{B}}_{l^\mathrm{s}_{\bar{n}},k} \right|^{2}+\sigma_{\mathrm{R}_{2}}^{2}\Vert\mathbf{h}_{\mathrm{B},l^\mathrm{s}_{\bar{n}}}\mathbf{A}\Vert^{2}+\sigma_{\mathrm{B}}^{2}}.
\label{eq:omega}
\end{aligned}
\end{equation}

\begin{figure*}[hbp]
	\setcounter{equation}{45}
	\hrulefill
	\begin{equation}
		\begin{scriptsize}
			\begin{aligned}
				&\nabla_{\boldsymbol{\theta}_{1,i}} \Phi_{1}\!=\hspace{-1mm}\sum\limits_{l^\mathrm{s}_{\bar{n}}\in \mathcal{L}_\mathrm{s}} \hspace{-1mm}\dfrac{2}{\ln2}  \left(-\dfrac{\sum\limits_{k\in \mathcal{L}_\mathrm{s}}\hspace{-1mm} p_{k} \boldsymbol{\mu}_{l^\mathrm{s}_{\bar{n}}\!,k}\boldsymbol{\mu}_{l^\mathrm{s}_{\bar{n}}\!,k}^\mathrm{H}\boldsymbol{\theta}_{1,i} }{\sum\limits_{ k\in \mathcal{L}_\mathrm{s}}\hspace{-1mm} p_{k}\hspace{-0.5mm}\left| \boldsymbol{\theta}_{1,i}^\mathrm{H} \boldsymbol{\mu}_{l^\mathrm{s}_{\bar{n}},k} \right|^{2}+\sigma_{\mathrm{R}_{2}}^{2}\Vert\mathbf{h}_{\mathrm{B},l^\mathrm{s}_{\bar{n}}}\mathbf{A}\Vert^{2}+\sigma_{\mathrm{B}}^{2}}+\dfrac{\sum\limits_{k\neq l^\mathrm{s}_{\bar{n}}\!,k\in \mathcal{L}_\mathrm{s}} \hspace{-4mm}p_{k} \boldsymbol{\mu}_{l^\mathrm{s}_{\bar{n}}\!,k}\boldsymbol{\mu}_{l^\mathrm{s}_{\bar{n}}\!,k}^\mathrm{H}\boldsymbol{\theta}_{1,i}}{\sum\limits_{k\neq l^\mathrm{s}_{\bar{n}},k\in \mathcal{L}_\mathrm{s}}\hspace{-4mm} p_{k}\hspace{-0.5mm}\left| \boldsymbol{\theta}_{1,i}^\mathrm{H} \boldsymbol{\mu}_{l^\mathrm{s}_{\bar{n}},k} \right|^{2}+\sigma_{\mathrm{R}_{2}}^{2}\Vert\mathbf{h}_{\mathrm{B},l^\mathrm{s}_{\bar{n}}}\mathbf{A}\Vert^{2}+\sigma_{\mathrm{B}}^{2}} \right.\\
				& \left.+\dfrac{\sum\limits_{k\in \mathcal{L}_\mathrm{s}} \hspace{-2mm}p_{k}\hspace{-1mm}\left(\hspace{-1mm}\boldsymbol{\eta}_{l^\mathrm{s}_{\bar{n}}\hspace{-0.3mm},\hspace{-0.2mm}k}\hspace{-0.3mm}\boldsymbol{\eta}_{l^\mathrm{s}_{\bar{n}}\hspace{-0.3mm},\hspace{-0.2mm}k}^\mathrm{H}\boldsymbol{\theta}_{\hspace{-0.3mm}1\hspace{-0.3mm},\hspace{-0.3mm}i}\hspace{-1mm}+\hspace{-0.5mm}\boldsymbol{\eta}_{l^\mathrm{s}_{\bar{n}}\hspace{-0.2mm},\hspace{-0.2mm}k}\hspace{-0.3mm}\zeta_{l^\mathrm{s}_{\bar{n}}\!,k}^{*}\hspace{-0.7mm}\right)\hspace{-1mm}+\hspace{-0.5mm}\sigma_\mathrm{\ddot{z}}^2\hspace{-2mm}\sum\limits_{l^\mathrm{z}_{\ddot{n}}\!\in \mathcal{L}_\mathrm{\ddot{z}}}\hspace{-3mm}\left(\hspace{-1mm}\boldsymbol{\eta}_{l^\mathrm{s}_{\bar{n}}\hspace{-0.2mm},\hspace{-0.2mm}l^\mathrm{z}_{\ddot{n}}}\hspace{-0.3mm}\boldsymbol{\eta}_{l^\mathrm{s}_{\bar{n}}\hspace{-0.3mm},\hspace{-0.2mm}l^\mathrm{z}_{\ddot{n}}}^\mathrm{H}\hspace{-0.3mm}\boldsymbol{\theta}_{\hspace{-0.3mm}1\hspace{-0.3mm},\hspace{-0.3mm}i}\hspace{-0.5mm}+\hspace{-0.5mm}\boldsymbol{\eta}_{l^\mathrm{s}_{\bar{n}}\hspace{-0.3mm},\hspace{-0.2mm}l^\mathrm{z}_{\ddot{n}}}\hspace{-0.4mm}\zeta_{l^\mathrm{s}_{\bar{n}}\hspace{-0.3mm},\hspace{-0.2mm}l^\mathrm{z}_{\ddot{n}}}^{*}\hspace{-0.5mm} \right) }{\sum\limits_{k\in \mathcal{L}_\mathrm{s}} \hspace{-1.5mm}p_{k}\hspace{-0.5mm}\left|  \zeta_{l^\mathrm{s}_{\bar{n}},k}\hspace{-0.5mm}+\hspace{-0.5mm} \boldsymbol{\theta}_{1,i}^\mathrm{H} \boldsymbol{\eta}_{l^\mathrm{s}_{\bar{n}},k}\hspace{-0.5mm} \right|^{2}\hspace{-0.5mm}+\sigma_\mathrm{\ddot{z}}^2\hspace{-1mm}\sum\limits_{ l^\mathrm{z}_{\ddot{n}} \!\in \mathcal{L}_\mathrm{\ddot{z}}}\!\! \left|  \zeta_{l^\mathrm{s}_{\bar{n}}\!,l^\mathrm{z}_{\ddot{n}}}+ \boldsymbol{\theta}_{1,i}^\mathrm{H} \boldsymbol{\eta}_{l^\mathrm{s}_{\bar{n}}\!,l^\mathrm{z}_{\ddot{n}}}\! \right|^{2}\!+\!\sigma_{\mathrm{E}}^{2}}\!-\!\dfrac{\sum\limits_{k\neq l^\mathrm{s}_{\bar{n}}\!, k\in \mathcal{L}_\mathrm{s}} \hspace{-5mm}p_{k}\hspace{-1mm}\left(\hspace{-1mm}\boldsymbol{\eta}_{l^\mathrm{s}_{\bar{n}}\hspace{-0.3mm},\hspace{-0.2mm}k}\boldsymbol{\eta}_{l^\mathrm{s}_{\bar{n}}\hspace{-0.3mm},\hspace{-0.2mm}k}^\mathrm{H}\hspace{-0.2mm}\boldsymbol{\theta}_{\hspace{-0.3mm}1\hspace{-0.3mm},\hspace{-0.3mm}i}\hspace{-0.7mm}+\hspace{-0.7mm}\boldsymbol{\eta}_{l^\mathrm{s}_{\bar{n}}\hspace{-0.3mm},\hspace{-0.2mm}k}\zeta_{l^\mathrm{s}_{\bar{n}}\hspace{-0.3mm},\hspace{-0.2mm}k}^{*}\hspace{-1mm}\right)\hspace{-0.7mm}+\hspace{-0.7mm}\sigma_\mathrm{\ddot{z}}^2\hspace{-1mm}\sum\limits_{l^\mathrm{z}_{\ddot{n}}\in \mathcal{L}_\mathrm{\ddot{z}}}\hspace{-3mm}\left(\hspace{-1mm}\boldsymbol{\eta}_{l^\mathrm{s}_{\bar{n}}\hspace{-0.3mm},\hspace{-0.2mm}l^\mathrm{z}_{\ddot{n}}}\hspace{-0.5mm}\boldsymbol{\eta}_{l^\mathrm{s}_{\bar{n}}\hspace{-0.3mm},\hspace{-0.2mm}l^\mathrm{z}_{\ddot{n}}}^\mathrm{H}\hspace{-0.5mm}\boldsymbol{\theta}_{\hspace{-0.3mm}1\hspace{-0.3mm},\hspace{-0.3mm}i}\hspace{-0.7mm}+\hspace{-0.7mm}\boldsymbol{\eta}_{l^\mathrm{s}_{\bar{n}}\hspace{-0.3mm},\hspace{-0.2mm}l^\mathrm{z}_{\ddot{n}}}\hspace{-0.5mm}\zeta_{l^\mathrm{s}_{\bar{n}}\hspace{-0.3mm},\hspace{-0.2mm}l^\mathrm{z}_{\ddot{n}}}^{*} \hspace{-0.5mm}\right) }{\sum\limits_{k\neq l^\mathrm{s}_{\bar{n}},k\in \mathcal{L}_\mathrm{s}} \hspace{-4mm}p_{k}\!\left|  \zeta_{l^\mathrm{s}_{\bar{n}},k}\hspace{-0.5mm}+\hspace{-0.5mm} \boldsymbol{\theta}_{1,i}^\mathrm{H} \boldsymbol{\eta}_{l^\mathrm{s}_{\bar{n}},k} \right|^{2}\hspace{-0.5mm}+\hspace{-0.5mm}\sigma_\mathrm{\ddot{z}}^2\!\hspace{-0.5mm}\sum\limits_{ l^\mathrm{z}_{\ddot{n}} \!\in \mathcal{L}_\mathrm{\ddot{z}}}\!\! \hspace{-1mm}\left|  \zeta_{l^\mathrm{s}_{\bar{n}}\!,l^\mathrm{z}_{\ddot{n}}}+ \boldsymbol{\theta}_{1,i}^\mathrm{H} \boldsymbol{\eta}_{l^\mathrm{s}_{\bar{n}}\!,l^\mathrm{z}_{\ddot{n}}}\hspace{-0.5mm} \right|^{2}\hspace{-0.5mm}+\hspace{-0.5mm}\sigma_{\mathrm{E}}^{2}}\hspace{-1mm}\right)\hspace{-1mm}.
			\end{aligned}
		\end{scriptsize}
		\label{eq:Euclidean gradient1}
	\end{equation}
	
\end{figure*}

Next, defining $\boldsymbol{\chi}_{l^\mathrm{s}_{\bar{n}},k}\hspace{-0.7mm}=\hspace{-0.7mm}\operatorname{diag} \hspace{-0.5mm}\left( \hspace{-1mm} \mathbf{f}_{l^\mathrm{s}_{\bar{n}}}^\mathrm{H}\mathbf{H}_{\mathrm{R_{2}B}} \boldsymbol{\Theta}_{2}\hspace{-1mm}\right)\hspace{-1mm} \mathbf{H}_{\mathrm{R_{1}}\mathrm{R_{2}}}\hspace{-0.5mm}\boldsymbol{\Theta}_{1}\hspace{-0.5mm} \mathbf{H}_{\mathrm{AR_{1}}}\hspace{-0.5mm}  \mathbf{f}_{k}  $ and $\boldsymbol{\kappa}_{l^\mathrm{s}_{\bar{n}}}\hspace{-0.7mm}=\hspace{-0.7mm}\operatorname{diag} \hspace{-0.5mm}\left( \hspace{-1mm} \mathbf{f}_{l^\mathrm{s}_{\bar{n}}}^\mathrm{H}\mathbf{H}_{\mathrm{R_{2}B}} \boldsymbol{\Theta}_{2}\hspace{-1mm}\right)$, we can rewrite $f_{\mathrm{B},l^\mathrm{s}_{\bar{n}}}$ in Eq.~\eqref{eq:f_B} as follows:
\setcounter{equation}{35}
\begin{equation}
	\sum\limits_{l^\mathrm{s}_{\bar{n}}\in \mathcal{L}_\mathrm{s}}f_{\mathrm{B},l^\mathrm{s}_{\bar{n}}}=-\mathbf{a}^\mathrm{H} \boldsymbol{\Omega} \mathbf{a}+2\Re\left\lbrace
	\mathbf{a}^\mathrm{H} \mathbf{g} \right\rbrace + c_{3},
\end{equation}	
where $c_{3}$ is a constant irrelevant
of $\mathbf{a}$, $\boldsymbol{\Omega} $ and $\mathbf{g}$ are respectively expressed by
\begin{equation}
 \boldsymbol{\Omega}=\sum\limits_{l^\mathrm{s}_{\bar{n}}\in \mathcal{L}_\mathrm{s}}\hspace{-1mm}\omega_{l^\mathrm{s}_{\bar{n}}}\hspace{-0.5mm}\left| \tau_{l^\mathrm{s}_{\bar{n}}}\right|^{2}\hspace{-1mm}\left(\sum\limits_{k\in \mathcal{L}_\mathrm{s}} p_{k} \boldsymbol{\chi}_{l^\mathrm{s}_{\bar{n}},k}\boldsymbol{\chi}_{l^\mathrm{s}_{\bar{n}},k}^\mathrm{H}+\sigma_{\mathrm{R}_{2}}^{2}\boldsymbol{\kappa_{l^\mathrm{s}_{\bar{n}}}}\boldsymbol{\kappa_{l^\mathrm{s}_{\bar{n}}}}^\mathrm{H} \hspace{-1mm}\right) 	
\end{equation}
and
\begin{equation}
	\mathbf{g}=\sum\limits_{l^\mathrm{s}_{\bar{n}}\in \mathcal{L}_\mathrm{s}}\omega_{l^\mathrm{s}_{\bar{n}}} \tau_{l^\mathrm{s}_{\bar{n}}}^{*}\sqrt{p_{l^\mathrm{s}_{\bar{n}}}} \boldsymbol{\chi}_{l^\mathrm{s}_{\bar{n}},l^\mathrm{s}_{\bar{n}}}.
\end{equation}

Based on the above analysis, we can obtain the optimal solution of  amplifying coefficients $\mathbf{a}$, denoted by $\mathbf{a}^{\star} $, by solving the following problem:
\setcounter{equation}{38}
\begin{subequations}
	\begin{align}
		\textbf{P5} : \; \max _{\mathbf{a} } & -\mathbf{a}^\mathrm{H} \boldsymbol{\Omega} \mathbf{a}+2\Re\left\lbrace
		\mathbf{a}^\mathrm{H} \mathbf{g} \right\rbrace + c_{3}  \\
		\mathrm{s.t}.\;
		& \mathbf{a}^\mathrm{H}
		\left( \sum\limits_{l^\mathrm{s}_{\bar{n}}\in \mathcal{L}_\mathrm{s}}\hspace{-1mm} p_{l^\mathrm{s}_{\bar{n}}} \boldsymbol{\nu}_{l^\mathrm{s}_{\bar{n}}}^\mathrm{H}\boldsymbol{\nu}_{l^\mathrm{s}_{\bar{n}}} \hspace{-0.5mm}+\hspace{-0.5mm} \sigma_\mathrm{\ddot{z}}^2\sum\limits_{l^\mathrm{z}_{\ddot{n}}\in \mathcal{L}_\mathrm{\ddot {z}}}\hspace{-1mm}\boldsymbol{\nu}_{l^\mathrm{z}_{\ddot{n}}}^\mathrm{H}\boldsymbol{\nu}_{l^\mathrm{z}_{\ddot{n}}}\hspace{-0.5mm}+\hspace{-0.5mm} \sigma_{\mathrm{R}_{2}}^{2} \mathbf{I}_{Q_{2}}\hspace{-1mm}\right)\hspace{-1mm} \mathbf{a} \nonumber \\
		& \leq P_{\mathrm{R}_{2}};\\
		&0\leq a_{q_{2}}\leq a_{\mathrm{max}},\forall q_{2} \in \left\lbrace 1,\dots, Q_{2}\right\rbrace,
	\end{align}
\end{subequations}
where $ \boldsymbol{\nu}_{\hat{l}}= \operatorname{diag}\left(\mathbf{H}_{\mathrm{R_{1}}\mathrm{R_{2}}}\hspace{-0.5mm}\boldsymbol{\Theta}_{1}\hspace{-0.5mm} \mathbf{H}_{\mathrm{AR_{1}}}\hspace{-0.5mm}  \mathbf{f}_{\hat{l}} \right) $ with $\hat{l} \in \left\lbrace l^\mathrm{s}_{\bar{n}},l^\mathrm{z}_{\ddot{n}}\right\rbrace $.
It is obvious that \textbf{P5} is a convex problem which can be solved by CVX.

\subsection{Optimize $ \boldsymbol{\theta}_{1} $ with given $ \mathbf{p}_\mathrm{s} $, $\mathbf{a}$, and $ \boldsymbol{\theta}_{2} $} \label{subsec:theta_1}
In the following, we calculate the optimal solution of $\boldsymbol{\theta}_{1}$, denoted by $\boldsymbol{\theta}_{1}^{\star}$, with given $\mathbf{p}_\mathrm{s}$, $\mathbf{a}$, and $\boldsymbol{\theta}_{2}$.

First, denoting $\zeta_{l^\mathrm{s}_{\bar{n}},\ddot{k}} =  \mathbf{h}_{\mathrm{AE},l^\mathrm{s}_{\bar{n}}} \mathbf{f}_{\ddot{k}}$, we respectively re-expressed $\gamma_{\mathrm{B}, l^\mathrm{s}_{\bar{n}}}$ and $\gamma_{\mathrm{E}, l^\mathrm{s}_{\bar{n}}}$ as follows:
\setcounter{equation}{39}
\begin{equation}
\left\{\hspace{-2mm}
\begin{array}{lr}
\gamma_{\mathrm{B}, l^\mathrm{s}_{\bar{n}}} \hspace{-1mm}=\hspace{-1mm} \frac{p_{l^\mathrm{s}_{\bar{n}}}\left| \boldsymbol{\theta}_{1}^\mathrm{H} \boldsymbol{\mu}_{l^\mathrm{s}_{\bar{n}},l^\mathrm{s}_{\bar{n}}} \right|^{2}}{\sum\limits_{ k \neq l^\mathrm{s}_{\bar{n}}, k \in \mathcal{L}_\mathrm{s}}\hspace{-4mm} p_{k}\left| \boldsymbol{\theta}_{1}^\mathrm{H} \boldsymbol{\mu}_{l^\mathrm{s}_{\bar{n}},k} \right|^{2}+\sigma_{\mathrm{R}_{2}}^{2}\Vert\mathbf{h}_{\mathrm{B},l^\mathrm{s}_{\bar{n}}}\mathbf{A}\Vert^{2}+\sigma_{\mathrm{B}}^{2}};\\
\gamma_{\mathrm{E}, l^\mathrm{s}_{\bar{n}}} \hspace{-1mm}=\hspace{-1mm} \frac{p_{l^\mathrm{s}_{\bar{n}}}\left|  \zeta_{l^\mathrm{s}_{\bar{n}},l^\mathrm{s}_{\bar{n}}}+ \boldsymbol{\theta}_{1}^\mathrm{H} \boldsymbol{\eta}_{l^\mathrm{s}_{\bar{n}}\!,l^\mathrm{s}_{\bar{n}}}  \right|^{2}}{\sum\limits_{ k \neq l^\mathrm{s}_{\bar{n}}, k \in \mathcal{L}_\mathrm{s}}\hspace{-4mm} p_{k}\left|  \zeta_{l^\mathrm{s}_{\bar{n}},k}\hspace{-0.5mm}+ \boldsymbol{\theta}_{1}^\mathrm{H} \boldsymbol{\eta}_{l^\mathrm{s}_{\bar{n}},k} \hspace{-0.5mm}\right|^{2}\hspace{-1mm}+\sigma_\mathrm{\ddot{z}}^2\hspace{-1mm}\sum\limits_{ l^\mathrm{z}_{\ddot{n}} \in \mathcal{L}_\mathrm{\ddot{z}}}\hspace{-1.5mm} \left|  \zeta_{l^\mathrm{s}_{\bar{n}},l^\mathrm{z}_{\ddot{n}}}\hspace{-0.5mm}+ \boldsymbol{\theta}_{1}^\mathrm{H} \boldsymbol{\eta}_{l^\mathrm{s}_{\bar{n}},l^\mathrm{z}_{\ddot{n}}} \right|^{2} \hspace{-1mm}+\sigma_{\mathrm{E}}^{2}},
\end{array}
\right.
\end{equation}
where $\boldsymbol{\mu}_{l^\mathrm{s}_{\bar{n}},k}$ and $\boldsymbol{\eta}_{l^\mathrm{s}_{\bar{n}},\ddot{k}}$ are respectively calculated by
\begin{equation}
\boldsymbol{\mu}_{l^\mathrm{s}_{\bar{n}},k} = \operatorname{diag}\left(\mathbf{f}_{l^\mathrm{s}_{\bar{n}}}^\mathrm{H} \mathbf{H}_{\mathrm{R_{2}B}}\boldsymbol{\Theta}_{2}\mathbf{A} \mathbf{H}_{\mathrm{R_{1}}\mathrm{R_{2}}}\right) \mathbf{H}_{\mathrm{AR_{1}}} \mathbf{f}_{k}
\end{equation}
and
\begin{equation}
\boldsymbol{\eta}_{l^\mathrm{s}_{\bar{n}},\ddot{k}} = \operatorname{diag}\left( \mathbf{h}_{\mathrm{R_{1}E},l^\mathrm{s}_{\bar{n}}}\right) \mathbf{H}_{\mathrm{AR_{1}}} \mathbf{f}_{\ddot{k}}.
\end{equation}
Thus, problem $\textbf{P1}$ can be re-formulated as follows:
\begin{subequations}
	\begin{align}
\textbf{P6}: \min _{\boldsymbol{\theta}_{1}}    &\;\Phi_{1}=-\left(R_{\mathrm{B}}-R_{\mathrm{E}}\right)\\ \quad\mathrm{s.t}. &\left|\theta_{1,q_{1}}\right|  = 1, \forall q_{1}\in \left\lbrace 1,\dots, Q\right\rbrace.
\end{align}
\end{subequations}
To solve problem $\textbf{P6}$ with given $ \mathbf{p}_\mathrm{s} $, $\bf{a}$, and $ \boldsymbol{\theta}_{2} $, the RMCG algorithm is required.
The unit modulus constraints $\left|\theta_{1,q_{1}}\right| = 1$ form a complex circle manifold $\mathcal{M}_{1} =\left\{\boldsymbol{\theta}_{1} \in \mathbb{C}^{Q_{1} \times 1}: \left|\theta_{1,1}\right| = \cdots =  \left|\theta_{1,Q_{1}}\right|= 1\right\}$. Thus, the search space of problem $ \textbf{P6} $ is the product of $Q_{1}$ complex circles, which in fact is Riemannian submanifold of $ \mathbb{C}^{Q_{1} \times 1} $. Thereby, we have the tangent space, denoted by $ T_{\boldsymbol{\theta}_{1,i}} \mathcal{M}_{1} $, for the $\boldsymbol{\theta}_{1,i}$ on the manifold $\mathcal{M}_{1}$  at the $i$-th iteration as follows:
\begin{equation}
T_{\boldsymbol{\theta}_{1,i}} \mathcal{M}_{1}=\left\{ \mathbf{v}_{1} \in \mathbb{C}^{Q_{1} \times 1} : \Re \left\{ \mathbf{v}_{1} \circ \boldsymbol{\theta}_{1,i}^{*} \right\}=\mathbf{0}_{Q_{1}}\right\},
\end{equation}
where $ \mathbf{v}_{1} $ denotes the tangent vector with respect to $\boldsymbol{\theta}_{1,i}$. The Riemannian gradient $ \operatorname{grad}_{\boldsymbol{\theta}_{1,i}} \Phi_{1} $ of the objective function $ \Phi_{1} $ in problem $\textbf{P6}$ at $\boldsymbol{\theta}_{1,i}$ is defined as the orthogonal projection of the Euclidean gradient $ \nabla_{\boldsymbol{\theta}_{1,i}} \Phi_{1}$ onto the tangent space $T_{\boldsymbol{\theta}_{1,i}} \mathcal{M}_{1}$. Thus,
\begin{equation}
\operatorname{grad}_{\boldsymbol{\theta}_{1,i}} \Phi_{1}=\nabla_{\boldsymbol{\theta}_{1,i}} \Phi_{1}-\Re \left\{\nabla_{\boldsymbol{\theta}_{1,i}} \Phi_{1} \circ \boldsymbol{\theta}_{1,i}^{*}  \right\} \circ \boldsymbol{\theta}_{1,i},
\label{eq:grad1}
\end{equation}
where $\nabla_{\boldsymbol{\theta}_{1,i}} f_{1}$ is derived as shown in Eq.~\eqref{eq:Euclidean gradient1}.

Then, we denote by $\alpha_{i}$ the Polak-Ribiere parameter \cite{Optimization_Algorithms_on_Matrix_Manifolds}, $\boldsymbol{\xi}_{i}$ the search direction at $\boldsymbol{\theta}_{1,i}$, and $\mathcal{T}_{\boldsymbol{\theta}_{1,i-1} \rightarrow \boldsymbol{\theta}_{1,i}}\left(\boldsymbol{\xi}_{i-1}\right)$ the vector transport function mapping a tangent vector from one tangent space to another tangent space. Thus, we can update the search direction $\boldsymbol{\xi}_{i}$ at the $i$-th iteration with the RMCG method as follows:
\setcounter{equation}{46}
\begin{equation}
	\begin{aligned}
\boldsymbol{\xi}_{i}=-\operatorname{grad}_{\boldsymbol{\theta}_{1,i}} \Phi_{1} +\alpha_{i}\mathcal{T}_{\boldsymbol{\theta}_{1,i-1} \rightarrow \boldsymbol{\theta}_{1,i}}\left(\boldsymbol{\xi}_{i-1}\right).
\end{aligned}
\label{eq:search direction1}
\end{equation}
Generally, $ \boldsymbol{\xi}_{i-1} $ and $ \boldsymbol{\xi}_{i} $ are respectively located in two different tangent spaces $ T_{\boldsymbol{\theta}_{1,i-1}} \mathcal{M} _{1}$ and $ T_{\boldsymbol{\theta}_{1,i}} \mathcal{M}_{1} $. Thereby, $ \mathcal{T}_{\boldsymbol{\theta}_{1,i-1} \rightarrow \boldsymbol{\theta}_{1,i}}\left(\boldsymbol{\xi}_{i}\right) $ can be derived as follows:
\begin{equation}
\begin{aligned}
\mathcal{T}_{\boldsymbol{\theta}_{1,i\!-\!1} \rightarrow \boldsymbol{\theta}_{1,i}}\hspace{-1mm}\left(\hspace{-0.5mm}\boldsymbol{\xi}_{i\!-\!1}\hspace{-0.5mm}\right) \hspace{-0.5mm}\triangleq\hspace{-0.5mm} T_{\boldsymbol{\theta}_{1,i\!-\!1}} \mathcal{M}_{1} &\hspace{-0.5mm}\mapsto\hspace{-0.5mm} T_{\boldsymbol{\theta}_{1,i}} \mathcal{M}_{1}:\\
\boldsymbol{\xi}_{i\!-\!1} &\hspace{-0.5mm}\mapsto\hspace{-0.5mm} \boldsymbol{\xi}_{i\!-\!1}\hspace{-0.5mm}-\hspace{-0.5mm}\Re\hspace{-0.5mm} \left\{\hspace{-0.5mm}\boldsymbol{\xi}_{i\!-\!1}\hspace{-0.5mm}\circ\hspace{-0.5mm}\boldsymbol{\theta}_{1,i}^{*}\hspace{-0.5mm}\right\}\hspace{-0.5mm}\circ\hspace{-0.5mm}\boldsymbol{\theta}_{1,i}.
\end{aligned}
\label{eq:vector transport1}
\end{equation}

According to $ \boldsymbol{\xi}_{i}$ and $\boldsymbol{\theta}_{1,i}$, $\boldsymbol{\theta}_{1,i+1}$ at the $(i+1)$-th iteration is updated by
\begin{equation}
\boldsymbol{\theta}_{1,i+1}=\operatorname{unt}\left(\boldsymbol{\theta}_{1,i}+\beta_{i}\boldsymbol{\xi}_{i} \right),
\label{eq:retraction1}
\end{equation}
where $\beta_{i}$ is the Armijo  line search step size and $\operatorname{unt}\left(\boldsymbol{\theta}_{1} \right)$ forms a vector whose elements are $\frac{\theta_{1,1}}{\vert \theta_{1,1}\vert},\ldots, \frac{\theta_{1,Q}}{\vert \theta_{1,Q}\vert}$.

Therefore, the optimal solution $\boldsymbol{\theta}_{1,i}^{\star}$ is obtained until the Riemannian gradient of the objective function $\operatorname{grad}_{\boldsymbol{\theta}_{1,i}} \Phi_{1}$ is close to zero~\cite{Optimization_Algorithms_on_Matrix_Manifolds}.

\begin{figure*}[hbp]
\setcounter{equation}{56}
	\hrulefill
  \begin{equation}
\begin{aligned}
	&\nabla_{\boldsymbol{\theta}_{2,i}} \Phi_{2}\!=\hspace{-2mm}\sum\limits_{l^\mathrm{s}_{\bar{n}}\in \mathcal{L}_\mathrm{s}} \hspace{-1.5mm}\dfrac{2}{\ln2}\hspace{-1.5mm}  \left(\hspace{-1mm}-\dfrac{\sum\limits_{k\in \mathcal{L}_\mathrm{s}}\hspace{-1mm} p_{k} \tilde{\boldsymbol{\mu}}_{l^\mathrm{s}_{\bar{n}}\!,k}\tilde{\boldsymbol{\mu}}_{l^\mathrm{s}_{\bar{n}}\!,k}^\mathrm{H}\boldsymbol{\theta}_{2,i} \hspace{-0.5mm}+\hspace{-0.5mm} \sigma_{\mathrm{R}_{2}}^{2}\boldsymbol{\iota}_{l^\mathrm{s}_{\bar{n}}}\boldsymbol{\iota}_{l^\mathrm{s}_{\bar{n}}}^\mathrm{H}\boldsymbol{\theta}_{2,i}}{\sum\limits_{ k\in \mathcal{L}_\mathrm{s}}\hspace{-1mm} p_{k}\hspace{-0.6mm}\left| \boldsymbol{\theta}_{2,i}^\mathrm{H} \tilde{\boldsymbol{\mu}}_{l^\mathrm{s}_{\bar{n}},k} \right|^{2}\hspace{-1.5mm}+\hspace{-0.5mm}\sigma_{\mathrm{R}_{2}}^{2}\Vert\boldsymbol{\theta}_{2,i}^\mathrm{H}\boldsymbol{\iota}_{l^\mathrm{s}_{\bar{n}}}\Vert^{2}\hspace{-1mm}+\hspace{-0.5mm}\sigma_{\mathrm{B}}^{2}}
	+\dfrac{\sum\limits_{k\neq l^\mathrm{s}_{\bar{n}}\!,k\in \mathcal{L}_\mathrm{s}} \hspace{-4mm}p_{k} \tilde{\boldsymbol{\mu}}_{l^\mathrm{s}_{\bar{n}}\!,k}\tilde{\boldsymbol{\mu}}_{l^\mathrm{s}_{\bar{n}}\!,k}^\mathrm{H}\boldsymbol{\theta}_{2,i}+\sigma_{\mathrm{R}_{2}}^{2}\boldsymbol{\iota}_{l^\mathrm{s}_{\bar{n}}}\boldsymbol{\iota}_{l^\mathrm{s}_{\bar{n}}}^\mathrm{H}\boldsymbol{\theta}_{2,i}}{\sum\limits_{k\neq l^\mathrm{s}_{\bar{n}},k\in \mathcal{L}_\mathrm{s}}\hspace{-1mm} p_{k}\hspace{-0.5mm}\left| \boldsymbol{\theta}_{2,i}^\mathrm{H} \tilde{\boldsymbol{\mu}}_{l^\mathrm{s}_{\bar{n}},k} \right|^{2}+\sigma_{\mathrm{R}_{2}}^{2}\Vert\boldsymbol{\theta}_{2,i}^\mathrm{H}\boldsymbol{\iota}_{l^\mathrm{s}_{\bar{n}}}\Vert^{2}+\sigma_{\mathrm{B}}^{2}}\right).
\label{eq:Euclidean gradient2}
\end{aligned}
\end{equation}
\end{figure*}

\subsection{Optimize $ \boldsymbol{\theta}_{2} $ with given $ \mathbf{p}_\mathrm{s} $, $\mathbf{a}$, and $ \boldsymbol{\theta}_{1} $}
In this subsection, the optimal solution of $\boldsymbol{\theta}_{2}$, denoted by $\boldsymbol{\theta}_{2}^{\star}$, with the given $\mathbf{p}_\mathrm{s}$, $\mathbf{a}$, and $\boldsymbol{\theta}_{1}$ also can be obtained by the RMCG algorithm. To calculate the maximum secrecy rate of OAM secure wireless communications, $\gamma_{\mathrm{B}, l^\mathrm{s}_{\bar{n}}}$ corresponding to $R_{B}$  is required to be re-expressed as follows:
\setcounter{equation}{49}
\begin{equation}	
		\gamma_{\mathrm{B}, l^\mathrm{s}_{\bar{n}}} \hspace{-1mm}=\hspace{-1mm} \frac{p_{l^\mathrm{s}_{\bar{n}}}\left| \boldsymbol{\theta}_{2}^\mathrm{H} \tilde{\boldsymbol{\mu}}_{l^\mathrm{s}_{\bar{n}},l^\mathrm{s}_{\bar{n}}} \right|^{2}}{\sum\limits_{ k \neq l^\mathrm{s}_{\bar{n}}, k \in \mathcal{L}_\mathrm{s}}\hspace{-4mm} p_{k}\left| \boldsymbol{\theta}_{2}^\mathrm{H} \tilde{\boldsymbol{\mu}}_{l^\mathrm{s}_{\bar{n}},k} \right|^{2}+\sigma_{\mathrm{R}_{2}}^{2}\Vert\boldsymbol{\theta}_{2}^\mathrm{H}\boldsymbol{\iota}_{l^\mathrm{s}_{\bar{n}}}\Vert^{2}+\sigma_{\mathrm{B}}^{2}},
\label{eq:theta_2}
\end{equation}
where $\tilde{\boldsymbol{\mu}}_{l^\mathrm{s}_{\bar{n}},k}$ and $\boldsymbol{\iota}_{l^\mathrm{s}_{\bar{n}}}$ are derived by
\begin{equation}
\begin{cases}
\tilde{\boldsymbol{\mu}}_{l^\mathrm{s}_{\bar{n}},k} = \operatorname{diag}\left(\mathbf{f}_{l^\mathrm{s}_{\bar{n}}}^\mathrm{H} \mathbf{H}_{\mathrm{R_{2}B}}\right)\mathbf{A} \mathbf{H}_{\mathrm{R_{1}}\mathrm{R_{2}}}\boldsymbol{\Theta}_{1} \mathbf{H}_{\mathrm{AR_{1}}} \mathbf{f}_{k};
\\
\boldsymbol{\iota}_{l^\mathrm{s}_{\bar{n}}} = \operatorname{diag}\left(\mathbf{f}_{l^\mathrm{s}_{\bar{n}}}^\mathrm{H} \mathbf{H}_{\mathrm{R_{2}B}}\right) \mathbf{A}.
\end{cases}
\end{equation}
Hence, substituting $\gamma_{\mathrm{B}, l^\mathrm{s}_{\bar{n}}}$ into Eq.~\eqref{eq:CB}, we can re-formulate problem $ \textbf{P1} $ as follows:
\begin{subequations}
	\begin{align}
		\textbf{P7}: \min _{\boldsymbol{\theta}_{2}}    &\;\Phi_{2}=-R_{\mathrm{B}}\\ \quad\mathrm{s.t}. &\left|\theta_{2,q_{2}}\right|  = 1, \forall q_{2}\in \left\lbrace 1,\dots, Q\right\rbrace.
	\end{align}
\end{subequations}
Similar to the analysis in subsection~\ref{subsec:theta_1}, the complex circle manifold $ \mathcal{M}_{2} =\left\{\boldsymbol{\theta}_{2} \in \mathbb{C}^{Q_{2} \times 1}: \left|\theta_{2,1}\right| = \cdots =  \left|\theta_{2,Q_{2}}\right|= 1\right\}$ is formed. To solve problem $\textbf{P7}$, the tangent space, denoted by $T_{\boldsymbol{\theta}_{2,i}} \mathcal{M}_{2}$, for given $\boldsymbol{\theta}_{2,i}$ on the manifold $\mathcal{M}_{2}$ at the $i$-th iteration can be derived as follows:
\begin{equation}
	T_{\boldsymbol{\theta}_{2,i}} \mathcal{M}_{2}=\left\{ \mathbf{v}_{2} \in \mathbb{C}^{Q_{2} \times 1} : \Re \left\{ \mathbf{v}_{2} \circ \boldsymbol{\theta}_{2,i}^{*} \right\}=\mathbf{0}_{Q_{2}}\right\},
\end{equation}
where $\mathbf{v}_{2}$ is the tangent vector at $\boldsymbol{\theta}_{2,i}$. Thus, the search direction, denoted by $\tilde{\boldsymbol{\xi}}_{i}$, at $\boldsymbol{\theta}_{2,i}$ can be updated as follows:
\begin{equation}
		\tilde{\boldsymbol{\xi}}_{i}=-\operatorname{grad}_{\boldsymbol{\theta}_{2,i}} \Phi_{2} +\tilde{\alpha}_{i}\mathcal{T}_{\boldsymbol{\theta}_{2,i-1} \rightarrow \boldsymbol{\theta}_{2,i}}\left(\tilde{\boldsymbol{\xi}}_{i-1}\right),
\label{eq:search direction2}
\end{equation}
where $\operatorname{grad}_{\boldsymbol{\theta}_{2,i}} \Phi_{2}$ is the Riemannian gradient of the function $\Phi_{2}$ with respect to $\boldsymbol{\theta}_{2,i}$ on the tangent space $ T_{\boldsymbol{\theta}_{2,i}} \mathcal{M}_{2} $, $ \tilde{\alpha}_{i} $ represents the Polak-Ribiere parameter to calculate $\boldsymbol{\theta}_{2}^{\star}$, and $ \mathcal{T}_{\boldsymbol{\theta}_{2,i-1} \rightarrow \boldsymbol{\theta}_{2,i}}\left(\tilde{\boldsymbol{\xi}}_{i-1}\right) $ is the vector transport function with respect to $\tilde{\boldsymbol{\xi}}_{i-1}$ mapping from $T_{\boldsymbol{\theta}_{2,i-1}} \mathcal{M}_{2}$ to $T_{\boldsymbol{\theta}_{2,i}} \mathcal{M}_{2}$. In Eq.~\eqref{eq:search direction2}, we can derive $\operatorname{grad}_{\boldsymbol{\theta}_{2,i}} \Phi_{2}$ and $ \mathcal{T}_{\boldsymbol{\theta}_{2,i-1} \rightarrow \boldsymbol{\theta}_{2,i}}\left(\tilde{\boldsymbol{\xi}}_{i-1}\right) $ as follows:
\begin{equation}
	\operatorname{grad}_{\boldsymbol{\theta}_{2,i}} \Phi_{2}=\nabla_{\boldsymbol{\theta}_{2,i}} \Phi_{2}-\Re \left\{\nabla_{\boldsymbol{\theta}_{2,i}} \Phi_{2} \circ \boldsymbol{\theta}_{2,i}^{*}  \right\} \circ \boldsymbol{\theta}_{2,i}
 \label{eq:grad2}
\end{equation}
and
\begin{equation}
\begin{aligned}
		\mathcal{T}_{\boldsymbol{\theta}_{2,i\!-\!1} \rightarrow \boldsymbol{\theta}_{2,i}}\hspace{-1mm}\left(\hspace{--.5mm}\tilde{\boldsymbol{\xi}}_{i\!-\!1}\hspace{-0.5mm}\right) \hspace{-0.5mm}\triangleq \hspace{-0.5mm} T_{\boldsymbol{\theta}_{2,i\!-\!1}} \mathcal{M}_{2} &\hspace{-0.5mm}\mapsto\hspace{-0.5mm} T_{\boldsymbol{\theta}_{2,i}} \mathcal{M}_{2}:
\\
		\tilde{\boldsymbol{\xi}}_{i\!-\!1} &\hspace{-0.5mm}\mapsto\hspace{-0.5mm} \tilde{\boldsymbol{\xi}}_{i\!-\!1}\hspace{-0.5mm}-\hspace{-0.5mm}\Re\hspace{-0.7mm} \left\{\hspace{-0.7mm}\tilde{\boldsymbol{\xi}}_{i\!-\!1}\hspace{-0.7mm}\circ\hspace{-0.7mm}\boldsymbol{\theta}_{2,i}^{*}\hspace{-0.7mm}\right\}\hspace{-0.7mm}\circ\hspace{-0.7mm}\boldsymbol{\theta}_{2,i},
	\end{aligned}
\label{eq:vector transport2}
\end{equation}
where $ \nabla_{\boldsymbol{\theta}_{2,i}} f_{2} $ can be derived as shown in Eq.~\eqref{eq:Euclidean gradient2}.

Then, denoting $ \tilde{\beta}_{i} $ the Armijo line search step size for $\boldsymbol{\theta}_{2,i}$, $\boldsymbol{\theta}_{2,i+1} $ at the $(i+1)$-th iteration can be updated as follows:
\setcounter{equation}{57}
\begin{equation}
		\boldsymbol{\theta}_{2,i+1}=\operatorname{unt}\left(\boldsymbol{\theta}_{2,i}+\tilde{\beta}_{i}\tilde{\boldsymbol{\xi}}_{i} \right).
\label{eq:retraction2}		
\end{equation}
The optimal $ \boldsymbol{\theta}_{2,i}^{\star} $ is obtained when the Riemannian gradient of the objective function is close to zero.

Based on the above-mentioned analysis, the entire algorithm for solving problem $ \textbf{P1} $ is summarized in Algorithm 1.

\subsection{Complexity Analysis}
In this subsection, we analyze the complexity of the overall alternative optimization algorithm.
Specifically, the computational complexities of optimizing $\mathbf{p}_\mathrm{s}$ in problem \textbf{P4} and optimizing $\mathbf{a}$ in problem \textbf{P5} are $ \mathcal{O}\left(N_\mathrm{s}^{3.5} \right) $ and $\mathcal{O}\left(Q_{2}^{3.5} \right)$, respectively.
The complexities of optimizing  $\boldsymbol{\theta}_{1}$ and $\boldsymbol{\theta}_{2}$ with the RMCG method mainly lie in calculating the Euclidean gradient in Eq.~\eqref{eq:Euclidean gradient1} and \eqref{eq:Euclidean gradient2}, which are $\mathcal{O}\left(N_\mathrm{s}^{2}Q_{1}^{2}\right) $ and $\mathcal{O}\left(N_\mathrm{s}^{2}Q_{2}^{2}\right) $, respectively. Thus, the total complexities of optimizing   $\boldsymbol{\theta}_{1}$ and $\boldsymbol{\theta}_{2}$ are $\mathcal{O}\left(L_{1}N_\mathrm{s}^{2}Q_{1}^{2}\right) $ and $\mathcal{O}\left(L_{2}N_\mathrm{s}^{2}Q_{2}^{2}\right) $, respectively, where $ L_{1}$ and $ L_{2}$ denote the iteration numbers for the convergence of the optimizations for $\boldsymbol{\theta}_{1}$ and $\boldsymbol{\theta}_{2}$, respectively. Therefore, the whole complexity of solving problem \textbf{P1} is $\mathcal{O}\left(L_{3} \left(N_\mathrm{s}^{3.5}+Q_{2}^{3.5}+ L_{1}N_\mathrm{s}^{2}Q_{1}^{2}+L_{2}N_\mathrm{s}^{2}Q_{2}^{2}\right)\right) $, where $L_{3} $ is the number of alternating optimization in Algorithm 1.

\begin{algorithm}[t]
	
	\caption{The RMCG Based Alternative Optimization}
	\label{alg:1}
	\begin{algorithmic}[1]
		\STATE Initialize $ \mathbf{p}_{\mathrm{s},0} $, $\mathbf{a}_{0}$, $ \boldsymbol{\theta}_{1,0} $, and $ \boldsymbol{\theta}_{2,0} $,
		calculate $ \boldsymbol{\xi}_{0}=-\operatorname{grad}_{\boldsymbol{\theta}_{1,0}}\Phi_{1} $ and $ \tilde{\boldsymbol{\xi}}_{0}=-\operatorname{grad}_{\boldsymbol{\theta}_{2,0}}\Phi_{2} $ according to Eqs.~\eqref{eq:grad1} and~\eqref{eq:grad2}, respectively, and set $ i=0 $;
		\STATE \textbf{repeat}
		\STATE With given $ \mathbf{p}_{\mathrm{s},i} $ at the $i$-th iteration, calculate $ t_{\mathrm{B},l^\mathrm{s}_{\bar{n}},i+1} $ and $ t_{\mathrm{E},l^\mathrm{s}_{\bar{n}},i+1} $ with Eq.~\eqref{eq:t_{B,E,l}};
		\STATE Given $ t_{\mathrm{B},l^\mathrm{s}_{\bar{n}},i+1} $ and $ t_{\mathrm{E},l^\mathrm{s}_{\bar{n}},i+1} $, calculate $ \mathbf{p}_{\mathrm{s},i+1} $ by solving $ \textbf{P4} $;
		
		\STATE Given $\mathbf{a}_{i} $, calculate $\tau_{l^\mathrm{s}_{\bar{n}},i+1}$ and $\omega_{l^\mathrm{s}_{\bar{n}},i+1}$ with Eqs.~\eqref{eq:tau} and \eqref{eq:omega}, respectively;
		\STATE Update $\mathbf{a}_{i+1} $ by solving \textbf{P5};
		
		\STATE Given $ \boldsymbol{\theta}_{1,i} $, calculate the Riemannian gradient $ \operatorname{grad}_{\boldsymbol{\theta}_{1,i}} \Phi_{1} $ with Eq.~\eqref{eq:grad1};
		\STATE Calculate the vector transport $\mathcal{T}_{\boldsymbol{\theta}_{1,i-1} \rightarrow \boldsymbol{\theta}_{1,i}}\left(\hspace{-0.5mm}\boldsymbol{\xi}_{i-1}\hspace{-0.5mm}\right)$ with Eq.~\eqref{eq:vector transport1};
		\STATE Choose Polak-Ribiere parameter $ \alpha_{i} $ and then calculate the search direction $ \boldsymbol{\xi}_{i} $ with Eq.~\eqref{eq:search direction1};
		\STATE  Choose Armijo  line search step size $ \beta_{i} $;
		\STATE Find $ \boldsymbol{\theta}_{1,i+1} $ according to Eq.~\eqref{eq:retraction1};
		
		\STATE Given $ \boldsymbol{\theta}_{2,i} $, calculate Riemannian gradient $ \operatorname{grad}_{\boldsymbol{\theta}_{2,i}} \hspace{-1mm}f_{2} $ with Eq.~\eqref{eq:grad2};
		
		\STATE Calculate the vector transport $\mathcal{T}_{\boldsymbol{\theta}_{2,i-1} \rightarrow \boldsymbol{\theta}_{2,i}}\left(\tilde{\boldsymbol{\xi}}_{i-1}\right)$ with Eq.~\eqref{eq:vector transport2};
		\STATE Choose Polak-Ribiere parameter $ \tilde{\alpha}_{i} $ and then calculate the search direction $ \tilde{\boldsymbol{\xi}}_{i} $ with Eq.~\eqref{eq:search direction2};
		\STATE Choose Armijo  line search step size $ \tilde{\beta}_{i} $;
		\STATE Find $ \boldsymbol{\theta}_{2,i+1} $ according to Eq.~\eqref{eq:retraction2});
		
		\STATE Update $ i=i+1 $;
		\STATE \textbf{until} convergence.
        \STATE Achieve the maximum secrecy rate with the obtained $ \mathbf{p}_\mathrm{s}^{\star} $, $\mathbf{a}^{\star}$, $ \boldsymbol{\theta}_{1}^{\star} $, and $ \boldsymbol{\theta}_{2}^{\star} $.
	\end{algorithmic}
\end{algorithm}

\section{Numerical Results}\label{sec:Results}
To evaluate the performance of our proposed double-RIS-assisted OAM secure scheme in the near field, we present several numerical results. Throughout the whole numerical simulation, we set $ N=N_{\mathrm{E}}=8 $, $ \mathbf{u}_{\mathrm{B}}=[0,0,40]^{\mathrm{T}} $ m, $ \mathbf{u}_{\mathrm{R_{1}}}=[2,0,0.3]^{\mathrm{T}} $ m, $ \mathbf{u}_{\mathrm{R_{2}}}=[1,0,39.7]^{\mathrm{T}} $ m,  $\theta=0 $, $\varphi=-\pi/20 $, $\vartheta_{\mathrm{A_{\mathrm{x}}}}=\vartheta_{\mathrm{B_{\mathrm{x}}}}=\vartheta_{\mathrm{R_{1,\mathrm{x}}}}=\vartheta_{\mathrm{R_{2,\mathrm{x}}}}=0 $, $\vartheta_{\mathrm{R_{1,\mathrm{y}}}}=\pi/10$, $\vartheta_{\mathrm{R_{2,\mathrm{y}}}}=-\pi/10$,  $\vartheta_{\mathrm{E_{\mathrm{x}}}}=\vartheta_{\mathrm{E_{\mathrm{y}}}}=-\pi/4 $, $ d_{\mathrm{y}}=d_{\mathrm{z}}=0.05 $ m, $ N_{\mathrm{A}}=4$, $ N_\mathrm{s}=N_\mathrm{\ddot{z}}=3$, $a_{\mathrm{max}}=10$, $\sigma_{\mathrm{B}}^{2}=\sigma_{\mathrm{E}}^{2}=-20$ dBm, and $ \sigma_{\mathrm{R}_{2}}^{2}=-60$ dBm. For our proposed system, we set the total power consumption $P_{\mathrm{total}}=P_{\mathrm{T}}+ P_{\mathrm{R}_{2}}$ with $P_{\mathrm{T}}=0.9\times P_{\mathrm{total}}$.

To verify the superiority of our proposed scheme in the near field, we compare the performance metrics among our proposed scheme and several other schemes, where phase shifts of RISs are required to be optimized. 1) Passive and active (PA) RIS-MIMO: A passive RIS and an active RIS are deployed near the legitimate transmitter and receiver, respectively. Several antennas emit AN and the remaining antennas with the power $\rho P_\mathrm{T}$ being optimized send desired signals in the PA RIS-MIMO secure communications. 2) PA RIS-OAM without AN: All available OAM modes with the optimal power allocation scheme are used for desired signal transmission. 3) PA RIS-OAM with random phases: The phases of PA RISs are generated randomly from $\left[ 0,2\pi\right] $. 4) Double passive (DP) RIS-OAM: The double passive  RISs are respectively deployed near the legitimate transmitter and receiver in DP RIS-OAM communications. 5) Single active (SA) RIS-OAM: The transmit power allocation $\rho P_\mathrm{T}$ for desired signal transmission is required to be jointly optimized with the phase shifts of RIS in the SA RIS-OAM system. 6) Single passive (SP) RIS-OAM: We jointly optimize the transmit power allocation and phase shifts in SP RIS-OAM system. 7) PA RIS-OAM with Zadoff-Chu (ZC) sequences: The DFT matrix for OAM generation is replaced by ZC sequences in PA RIS-OAM with ZC scheme as compared with our proposed scheme.



\begin{figure}
	\centering
	\vspace{0pt}
	\includegraphics[width=0.5\textwidth]{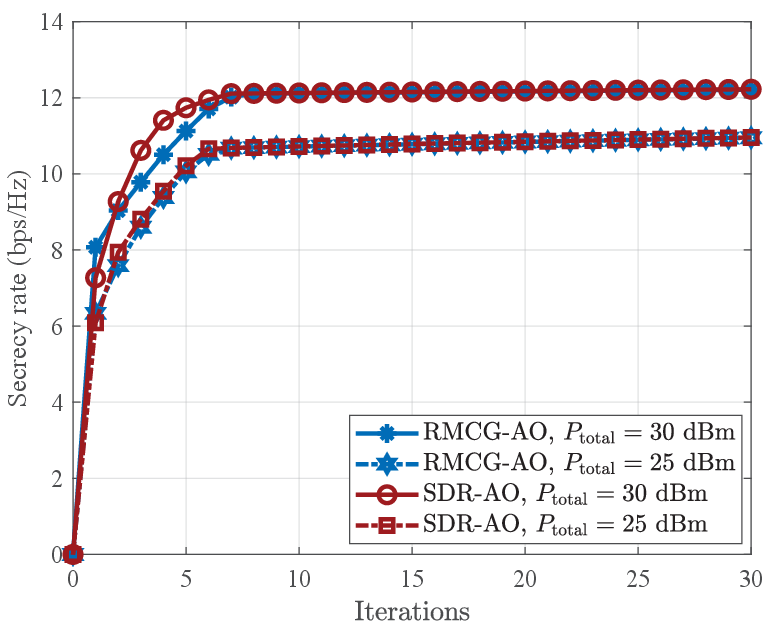}
	\caption{Secrecy rates versus the iterations.} \label{fig:iterations}
	\vspace{0pt}
\end{figure}

To verify the convergence of our developed RMCG-AO algorithm, we compare it with the semidefinite-relaxation-(SDR) based AO algorithm versus different  $P_\mathrm{total}$, where we set $Q_{1}=Q_{2}=40 $, $\vartheta_{\mathrm{A_{\mathrm{y}}}}=\pi/10$, $\vartheta_{\mathrm{B_{\mathrm{y}}}}=-\pi/10$, and $\rho=0.9$. It is obviously observed that the secrecy rates of these two algorithms are rapidly convergent to the maximum values as the iteration increases. The iteration number required by the RMCG-AO algorithm is close to the same as the SDR-AO algorithm, while the computational complexity of the former algorithm is lower. Therefore, Fig.~\ref{fig:iterations} demonstrates the superiority of our developed RMCG-AO algorithm.

\begin{figure}
	\centering
	\includegraphics[width=0.5\textwidth]{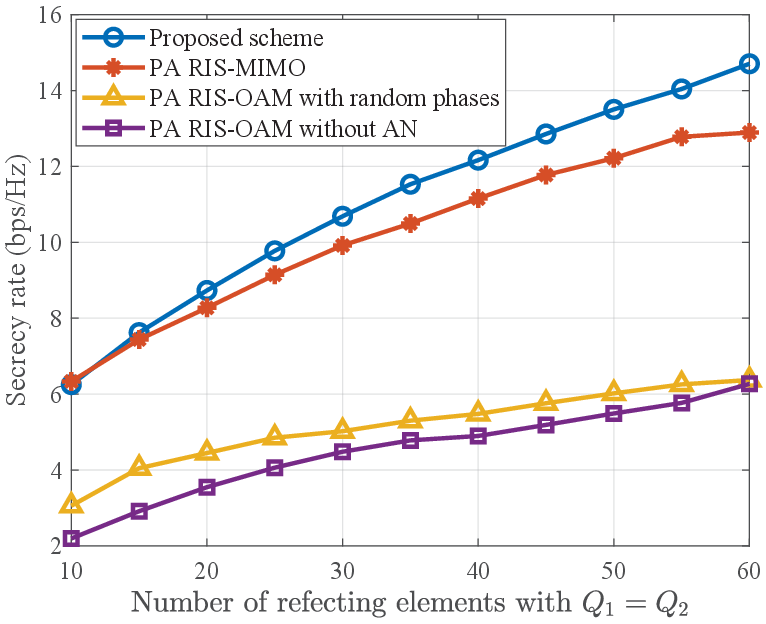}
	\caption{Secrecy rates versus the number of RIS reflecting elements with $Q_{1}=Q_{2}$.}
\label{fig:element}
\end{figure}

Figure \ref{fig:element} compares the secrecy rates of different schemes versus the number of RIS reflecting elements with $Q_{1}=Q_{2}$, where we set $\vartheta_{\mathrm{A_{\mathrm{y}}}}=\pi/10$, $ P_\mathrm{total}=30 $ dBm, $\rho=0.9$, and $\vartheta_{\mathrm{B_{\mathrm{y}}}}=-\pi/10$. Observed from Fig.~\ref{fig:element}, the increase of RIS reflecting elements brings to the increase of secrecy rates of all schemes. This happens because designing the RIS phase shifts can enhance the strength of desired signals received by the legitimate receivers while weakening the strength received by the eavesdropper, thus resulting in high secrecy rates in secure wireless communications. Due to the low DoFs in LoS MIMO secure communications and the hollow structure of OAM beams, our proposed double-RIS-assisted OAM secure scheme has higher secrecy rates than the conventional LoS MIMO secure communications with RISs. Since the PA RIS-OAM with random phases scheme cannot fully utilize the phase advantage of RIS, the secrecy rate increases slowly as $Q_{1}$ and $Q_{2}$ increase. 
In addition, the PA RIS-OAM without AN scheme performs worse than our proposed scheme, because the eavesdropper not jammed by AN has high achievable rates.


\begin{figure}
	\centering
	\includegraphics[width=0.5\textwidth]{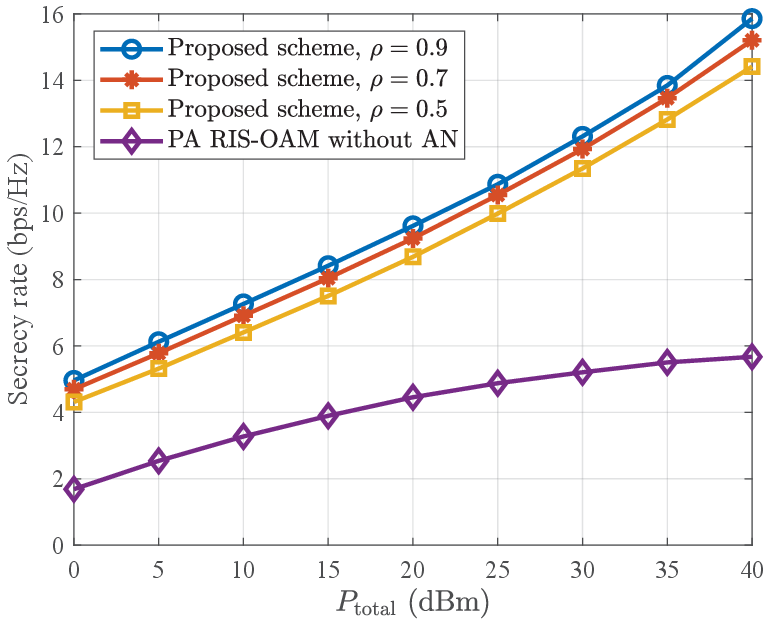}
	\caption{Secrecy rates versus the transmit power $P_{\mathrm{T}}$.} \label{fig:rho-Pt}
\end{figure}

Figure \ref{fig:rho-Pt} depicts the secrecy rates versus the total power consumption $P_\mathrm{total}$ for our proposed scheme and the PA RIS-OAM without AN scheme, where we set $Q_{1}=Q_{2}=40 $, $\vartheta_{\mathrm{A_{\mathrm{y}}}}=\pi/10$, $\vartheta_{\mathrm{B_{\mathrm{y}}}}=-\pi/10$, and $\rho=0.5, 0.7, 0.9$, respectively. As indicated in Fig.~\ref{fig:rho-Pt}, the secrecy rates of our proposed scheme increase as $\rho$ increases. This is explained by the fact that the transmit power for desired signals increases with the increase of $\rho$, thus resulting in a high achievable rate of Bob. Also, the achievable rate for Eve increases more slowly than that for Bob with the increase of $\rho$, thus achieving a high secrecy rate for our proposed scheme in the high $\rho$ region. Figs.~\ref{fig:element} and~\ref{fig:rho-Pt} demonstrate the effectiveness of employing AN in the near-field secure OAM communications.

\begin{figure}
	\centering
	\includegraphics[width=0.5\textwidth]{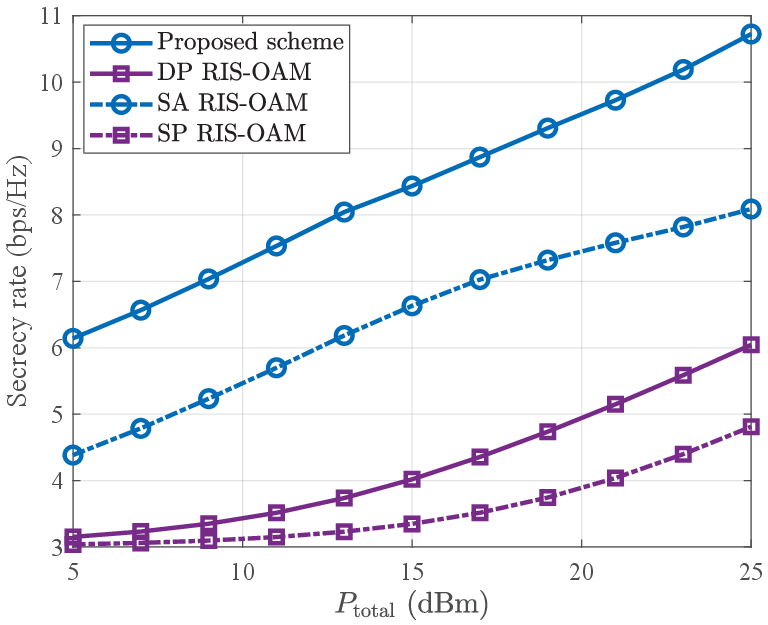}
	\caption{Secrecy rate comparisons between our proposed scheme and the single RIS-OAM scheme versus $\theta_{\mathrm{A_\mathrm{y}}}$.} \label{fig:thetaAy}
\end{figure}

Figure \ref{fig:thetaAy} indicates the secrecy rates versus  $P_\mathrm{total}$ for our proposed scheme, the DP RIS-OAM scheme, the SA RIS-OAM, and the SP RIS-OAM scheme, where we set   $\vartheta_{\mathrm{A_{\mathrm{y}}}}=\pi/10$, $\vartheta_{\mathrm{B_{\mathrm{y}}}}=-\pi/10$, and $\rho=0.9$. Also, we set the total number of RIS reflecting elements for these four schemes as 80. It is obvious that our proposed scheme (or the SA RIS-OAM scheme) is superior to the DP RIS-OAM scheme (or the SP RIS-OAM scheme) in term of secrecy rates with the same $P_\mathrm{total}$. This is because that active RIS can amplify the attenuated signals with a small amount of total power, thus overcoming the double-path fading effect and significantly enhancing the signal strength received by the legitimate receiver. In addition,
although these four schemes have the same number of RIS reflecting elements, the secrecy rate of our proposed scheme (or the DP RIS-OAM scheme) is higher than that of the SA RIS-OAM scheme (or the SP RIS-OAM scheme). The major reason is that our proposed scheme (or the DP RIS-OAM scheme) can adjust the OAM beam direction twice with the double RISs to dramatically mitigate the inter-mode interference caused by the misaligned UCA-based transceivers. In contrast, the OAM beam direction can be adjusted once with RIS in the SA RIS-OAM scheme (or the SP RIS-OAM scheme). Thus, the mitigation of inter-mode interference is restricted. Hence, Fig.~\ref{fig:thetaAy} proves the superiority of our proposed double-RIS-assisted OAM secure scheme.

\begin{figure}
	\centering
	\includegraphics[width=0.5\textwidth]{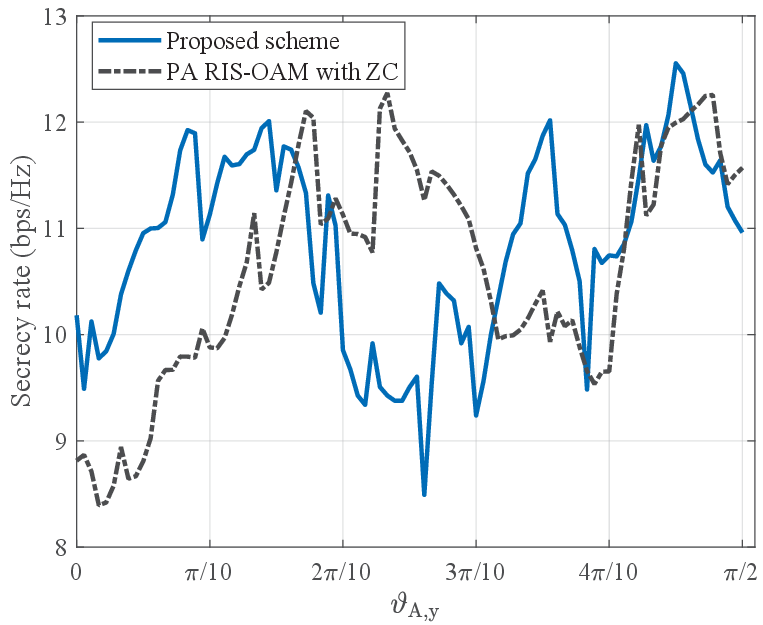}
	\caption{Secrecy rates for our proposed scheme and double RIS-OAM with ZC scheme versus $\theta_{\mathrm{A_\mathrm{y}}}$.} \label{fig:OAM-ZC}
\end{figure}
Figure \ref{fig:OAM-ZC} exhibits the secrecy rates versus $\theta_{\mathrm{A_\mathrm{y}}}$ for our proposed scheme and the PA RIS-OAM with ZC scheme, where $P_\mathrm{total}=30$ dBm, $Q_{1}=Q_{2}=40 $, $\vartheta_{\mathrm{B_{\mathrm{y}}}}=0$, and $\rho=0.9$. In the double RIS-OAM with ZC scheme, the ZC sequences are expressed by $e^{j\frac{\pi U(n-1)^2}{N}}$ if $N$ is an even number, where $U$ is an integer less than $N$. Then, the signal $\mathbf{x}$ can be re-expressed by
\begin{equation}
\mathbf{x} = \rm{diag} (\bf{w}) \bf{F} \tilde{\mathbf{s}},
\end{equation}
where $\rm{diag} (\bf{w})$ is the diagonal matrix composed of the ZC sequences $\bf{w}$~\cite{8761512}. At the receiver, the DFT algorithm is also performed to decompose the OAM signals in the PA RIS-OAM with ZC scheme. Thanks to the non-hollow beam structure and the convergence for OAM beams, the PA RIS-OAM with ZC scheme outperforms our proposed scheme in terms of the secrecy rate at around $ \pi/6\leq\theta_{\mathrm{A_\mathrm{y}}} \leq 3\pi/10$. This shows that the ZC sequences can be applied to significantly enhance the secrecy rate of near-field secure OAM communications when conventional OAM has poor security performance.

\begin{figure}
	\centering
	\includegraphics[width=0.5\textwidth]{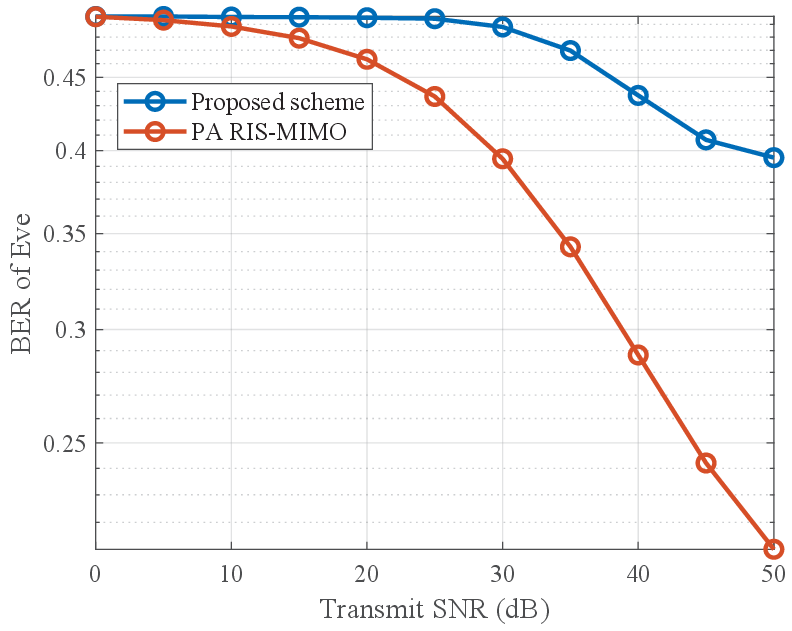}
	\caption{BERs of Eve versus the transmit SNR at Eve.} \label{fig:BER}
\end{figure}
Figure~\ref{fig:BER} illustrates the bit error rates (BERs) versus the transmit signal-to-noise ratio (SNR) for our proposed scheme and the PA RIS-MIMO scheme with quadrature phase shift keying modulation in the near field, where we set $ P_\mathrm{total}=30 $ dBm, $Q_{1}=Q_{2}=40 $,  $\vartheta_{\mathrm{A_{\mathrm{y}}}}=\pi/10$, $\vartheta_{\mathrm{B_{\mathrm{y}}}}=-\pi/10$, and $\rho=0.9$. As shown in Fig.~\ref{fig:BER}, the BER of Eve for our proposed scheme is greater than that for the PA RIS-MIMO scheme. On the one hand, the hollow structure and beam direction of OAM cause the fragmented information of desired signals sent by the legitimate transmitter. On the other hand, the Eve cannot easily decompose the received OAM desired signals especially when Eve uses the conventional signal recovery scheme in MIMO communications, thus causing severe phase crosstalk. As compared with our proposed scheme, the main impact on the BERs at the Eve with the PA RIS-MIMO scheme is the AN. Thereby, the Eve can more easily recover the desired signals with lower BERs with the PA RIS-MIMO scheme than our proposed scheme, thus wiretapping legitimate information. Figs.~\ref{fig:element} and~\ref{fig:BER} show that our proposed scheme has better anti-eavesdropping results as compared with the conventional PA RIS-MIMO scheme.

\section{Conclusions}\label{sec:Conclusion}
In this paper, we proposed the double-RIS-assisted OAM with JiMa secure near-field communication scheme, where double RIS are easily deployed to reconstruct the direct links between the legitimate transceivers and mitigate the inter-mode interference caused by the misaligned legitimate transceiver, to significantly increase the secrecy rate of wireless communications. The JiMa scheme, which jointly designs the OAM modes selection for desired signals and AN transmissions with index modulation, was proposed to weaken the desired signals strength received by eavesdroppers. Also, the RMCG-AO algorithm was developed to jointly optimize the transmit power allocation and phase shifts of double RISs, thus maximizing the secrecy rate subject to the total transmit power, unit modulus of RISs' phase shifts, and allocated power threshold. Numerical results have demonstrated that our proposed scheme is superiority of the existing works in terms of BERs and secrecy rates. In addition, it has been shown that the ZC sequences can be applied into our proposed scheme to significantly enhance the secrecy rate of near-field secure OAM communications when the conventional OAM shows poor security performance.

\begin{appendices}
\section{Theorem 1}\label{app:T1}
The first term on the right hand of Eq.~\eqref{eq:I_sy} can be derived as follows:
\begin{eqnarray}
    \mathcal{I}(\bm{s},\tilde{\mathbf{y}}_{\mathrm{B}}| \mathcal{L}_\mathrm{s},\mathcal{L_\mathrm{\ddot{z}}})=\sum\limits_{l^\mathrm{s}_{\bar{n}}\in \mathcal{L}_\mathrm{s}} \log _{2}\left(1+\gamma_{\mathrm{B}, l^\mathrm{s}_{\bar{n}}}\right).
\label{eq:I_s_im}
\end{eqnarray}

Since the received noise ${\bf n}_{\rm B}$ follows normal distribution, the decomposed signal $\tilde{\mathbf{y}}_{\mathrm{B}}$ also follows. Thus, the Gaussian mixture probability density function (PDF), denoted by $f(\tilde{\mathbf{y}}_{\mathrm{B}})$, considering all possible $G$ combinations is obtained as follows:
\begin{eqnarray}
    f(\tilde{\mathbf{y}}_{\mathrm{B}})=\frac{1}{G}\sum_{g=1}^{G}f(\tilde{\mathbf{y}}_{\mathrm{B},g}),
\end{eqnarray}
where $f(\tilde{\mathbf{y}}_{\mathrm{B},g})$ is the PDF of the received signals $\tilde{\mathbf{y}}_{\mathrm{B}}$ decomposed by Bob with the $g$-th SN pair. Thus, the expression of $\mathcal{I}(\mathcal{L}_\mathrm{s},\mathcal{L_\mathrm{\ddot{z}}},\tilde{\mathbf{y}}_{\mathrm{B}})$ can be written as the Kullback-Leibler (KL) divergence between the Gaussian distribution with PDF $f(\tilde{\mathbf{y}}_{\mathrm{B},g})$ and Gaussian mixture with the PDF $f(\tilde{\mathbf{y}}_{\mathrm{B}})$, which is given by
\begin{eqnarray}
    \mathcal{I}(\mathcal{L}_\mathrm{s},\mathcal{L_\mathrm{\ddot{z}}},\tilde{\mathbf{y}}_{\mathrm{B}}) = \frac{1}{G}\sum_{g=1}^{G} \mathcal{D}\left[f(\tilde{\mathbf{y}}_{\mathrm{B},g})\parallel f(\tilde{\mathbf{y}}_{\mathrm{B}})\right],
    \label{eq:I_lk}
\end{eqnarray}
where
\begin{eqnarray}
  &&\hspace{-0.7cm}\mathcal{D}\left[f(\tilde{\mathbf{y}}_{\mathrm{B},g})\parallel f(\tilde{\mathbf{y}}_{\mathrm{B}})\right]
  \nonumber\\
   &&\hspace{-0.7cm}=\int f(\tilde{\mathbf{y}}_{\mathrm{B},g})\log_{2} \frac{f(\tilde{\mathbf{y}}_{\mathrm{B},g})}{f(\tilde{\mathbf{y}}_{\mathrm{B}})} d \bm{\tilde{\mathbf{y}}}_{\mathrm{B}}
   \nonumber\\
   &&\hspace{-0.7cm}=  \int f(\tilde{\mathbf{y}}_{\mathrm{B},g})\log_{2}f(\tilde{\mathbf{y}}_{\mathrm{B},g})d \bm{\tilde{\mathbf{y}}}_{\mathrm{B}}-\int f(\tilde{\mathbf{y}}_{\mathrm{B},g})\log_{2}f(\tilde{\mathbf{y}}_{\mathrm{B}})d \bm{\tilde{\mathbf{y}}}_{\mathrm{B}}.
   \nonumber\\
   \label{eq:kl}
\end{eqnarray}
The second term on the right hand of Eq.~\eqref{eq:kl} is upper bounded as (\cite{KL_2008}, Eq. (9))
\begin{eqnarray}
   &&\hspace{-1.2cm}-\int f(\tilde{\mathbf{y}}_{\mathrm{B},g})\log_{2}f(\tilde{\mathbf{y}}_{\mathrm{B}})d \bm{\tilde{\mathbf{y}}}_{\mathrm{B}}
   \nonumber\\
  &&\hspace{-1.2cm}=- \int f(\tilde{\mathbf{y}}_{\mathrm{B},g})\log_{2} \left[\frac{1}{G}f(\tilde{\mathbf{y}}_{\mathrm{B},g})(1+\epsilon_{g})\right]d \bm{\tilde{\mathbf{y}}}_{\mathrm{B}}
  \nonumber\\
    &&\hspace{-1.2cm}=- \int f(\tilde{\mathbf{y}}_{\mathrm{B},g}) \left[\log_{2}f(\tilde{\mathbf{y}}_{\mathrm{B},g})-\log_{2}G+\log_{2}(1+\epsilon_{g})\right]d \bm{\tilde{\mathbf{y}}}_{\mathrm{B}},
    \nonumber\\
\end{eqnarray}
where
\begin{equation}
    \epsilon_{g}=\frac{\sum_{g\neq i \ i=1}^{G}f(\tilde{\mathbf{y}}_{\mathrm{B},i})}{f(\tilde{\mathbf{y}}_{\mathrm{B},g})}.
\end{equation}
Since $\epsilon_{g}$ is positive, $\log_{2}(1+\epsilon_{g})$ is always non-negative. Neglecting it, we still can obtain the upper bound. Thus, we have
\begin{equation}
   \int \!\!\! f(\tilde{\mathbf{y}}_{\mathrm{B},g})\!\log_{2}f(\tilde{\mathbf{y}}_{\mathrm{B}})d \bm{\tilde{\mathbf{y}}}_{\mathrm{B}} \!\! \geq \!\! \int \!\!\! f(\tilde{\mathbf{y}}_{\mathrm{B},g})\!\log_{2}f(\tilde{\mathbf{y}}_{\mathrm{B},g})d \bm{\tilde{\mathbf{y}}}_{\mathrm{B}}-\log_{2}\!{G}.
    \label{eq:kl_2}
\end{equation}
Combining Eqs.~\eqref{eq:kl} and \eqref{eq:kl_2}, we have
\begin{eqnarray}
    \mathcal{D}\left[f(\tilde{\mathbf{y}}_{\mathrm{B},g})\parallel f(\tilde{\mathbf{y}}_{\mathrm{B}})\right] \leq \log_{2}{G}.
\end{eqnarray}
Therefore, the upper bound of achievable rate at Bob is calculated as Eq.~\eqref{eq:CB}.
\end{appendices}

\bibliographystyle{IEEEtran}
\bibliography{References}

\end{document}